\documentclass[12pt]{iopart}

\expandafter\let\csname equation*\endcsname\relax
\expandafter\let\csname endequation*\endcsname\relax
\usepackage[utf8]{inputenc}                             
\usepackage[final,nopatch=footnote]{microtype}
\usepackage[T1]{fontenc}                                
\usepackage{hyperref}                                   
\usepackage{url}                                        
\usepackage{booktabs}                                   
\usepackage{amsfonts,amssymb,amsmath}                   
\usepackage{bm,graphicx}
\usepackage{microtype}                                  
\usepackage{setspace} 
\usepackage[colorinlistoftodos]{todonotes}                

\usepackage{etoolbox}
\usepackage[normalem]{ulem}

\providecolor{added}{rgb}{0,0,1}
\providecolor{deleted}{rgb}{1,0,0}

\makeatletter
\newrobustcmd{\fixappendix}{%
  \patchcmd{\l@section}{1.5em}{7em}{}{}%
  \patchcmd{\l@subsection}{2.3em}{7em}{}{}%
}
\makeatother

\usepackage[utf8]{inputenc}
\usepackage{amsmath}
\usepackage{amsfonts}
\usepackage{amssymb}
\usepackage{makeidx}
\usepackage{graphicx}
\usepackage{color}
\usepackage{mathptmx}
\usepackage{bm}
\usepackage{soul}
\usepackage[mathscr]{euscript}
\usepackage{tikz}
\usepackage{hyperref}
\usepackage[makeroom]{cancel}

\newcommand{\erefs}[1]{Eqs.~(\ref{#1})}

\def\bea{\begin{eqnarray}}
\def\eea{\end{eqnarray}}
\def\ba{\begin{array}}
\def\ea{\end{array}}

\def\la{\langle}
\def\ra{\rangle}

\begin{document}

\title{Stochastic Two-temperature Nonequilibrium Ising model}

\author{Debraj Dutta} 
\address{S.~N.~Bose National Centre for Basic Sciences, Kolkata 700106, India}
\ead{debraj.dutta@bose.res.in}

\author{Ritwick Sarkar}
\address{S.~N.~Bose National Centre for Basic Sciences, Kolkata 700106, India}
\ead{ritwick.sarkar@bose.res.in}

\author{Urna Basu\footnote{Corresponding author}}
\address{S.~N.~Bose National Centre for Basic Sciences, Kolkata 700106, India}
\ead{urna@bose.res.in}

\begin{abstract}
We investigate the nonequilibrium stationary state (NESS) of the two-dimensional Ising model under a stochastic dichotomous modulation of temperature, which alternates between $T_c \pm \delta$ around the critical temperature $T_c$ at a rate $\gamma$. Both magnetization and energy exhibit non-monotonic dependence on $\gamma$, explained by a renewal approach in the slow-switching limit, while for small $\delta$ dynamical response theory quantitatively captures the $\gamma$-dependence of the observables. In the fast-switching regime, the NESS appears Boltzmann-like with a $\gamma$-dependent effective temperature. However, a finite energy current flowing through the system from hot to cold reservoir confirms the intrinsic nonequilibrium nature of the dynamics.
\end{abstract}

\maketitle

\section{Introduction}\label{sec:intro}
The Ising model has served as a cornerstone of statistical physics for nearly a century~\cite{Ising_review, budrikis2024100}. Originally introduced to understand critical phenomena~\cite{ising1925beitrag}, it has been extensively studied using various dynamical rules~\cite{glauber_dynamics, kawasaki1966diffusion, kawasaki_2} that preserve detailed balance with respect to the equilibrium Boltzmann distribution. Despite its apparent simplicity, the Ising model exhibits a wide range of nontrivial phenomena, including phase transitions, domain coarsening, phase separation, and giant fluctuations. Owing to this rich behavior, its influence extends far beyond statistical physics, with applications in diverse fields such as biology, sociology, and economics~\cite{ising_biology_1, ising_socio_1, ising_economy_1}.

The Ising model also provides a natural framework for exploring out-of-equilibrium behaviour. Over the years, several nonequilibrium versions of the model have been explored in different contexts. A commonly studied setup involves coupling different regions of the system to thermal baths at different temperatures. This naturally drives currents and leads to nontrivial steady-state profiles of magnetization and energy. Such arrangements have been investigated in both one~\cite{zia2007probability, racz1994two, mazilu2009exact, lecomte2005energy, borchers2014nonequilibrium, lavrentovich2010energy, lavrentovich2012steady} and two dimensions~\cite{blote1990critical}. Generalisations involving spatially varying couplings or temperature fields have also been investigated~\cite{garrido1987ising,blote1991stability}. Another direction focuses on models that break detailed balance at the microscopic level—typically by introducing spin-flip rates that depend on two or more temperatures~\cite{Entropy_Production_2d_ising}. These systems have been examined using superstatistics~\cite{Najafi2021}, and comparisons across different update schemes (Metropolis, Glauber, Swendsen-Wang) suggesting that certain signatures of equilibrium criticality may persist even far from equilibrium~\cite{Tamayo1994}. In related studies, the interplay between finite and infinite temperature baths has been shown to induce interesting steady-state features, such as nonmonotonic variation of surface tension~\cite{PhysRevB.70.245409}. More recently, time-dependent temperature protocols have attracted attention, where the Ising system is alternately coupled to thermal baths at different temperatures. These studies range from exploring calorimetric response~\cite{Beyen2024} and entropy production rate~\cite{mamede2025} in mean-field models to exploring applications such as protein sequence inference~\cite{Dietler2025.05.14.654088}. However, the nonequilibrium stationary behaviour of the Ising model under such stochastic temporal driving remains mostly unexplored.

In this paper we investigate a scenario where the two-dimensional Ising model is subject to a stochastic dichotomous temperature. In particular, we consider the case where the temperature intermittently switches between two values $T_c \pm \delta$, with a constant rate $\gamma$. We characterize the resulting NESS by measuring the mean and fluctuations of magnetization and energy. Surprisingly, for large $\delta$, the average magnetization and energy both show a non-monotonic behaviour as $\gamma$ is increased. We show that this non-monotonicity originates from the difference in the typical relaxation times of the equilibrium Ising model below and above the critical temperature $T_c.$ On the other hand for small $\delta$, we characterise the NESS using dynamical response formalism and show that the second-order response coefficient changes its sign as $\gamma$ is increased. Numerical evidence further suggests that, in the large $\gamma$ regime, the stationary weight of the spin configurations admits a Boltzmann-like form with a $\gamma$ and $\delta$ dependent effective temperature $T_\text{eff} < T_c$. However the dynamics remains nonequilibrium, which is manifest from a persistent flow of energy current through the system.

The paper is organised as follows. In the next section we define the model and present a brief summary of our main results. Section~\ref{sec:NESS} focuses on the behaviour of magnetization and energy in the stationary state. The effective temperature picture in the large $\gamma$ regime is discussed in Sec.~\ref{sec:LG}. We conclude with some general remarks in Sec.~\ref{sec:conc}

\begin{figure}[t]
    \centering
    \includegraphics[width=14cm]{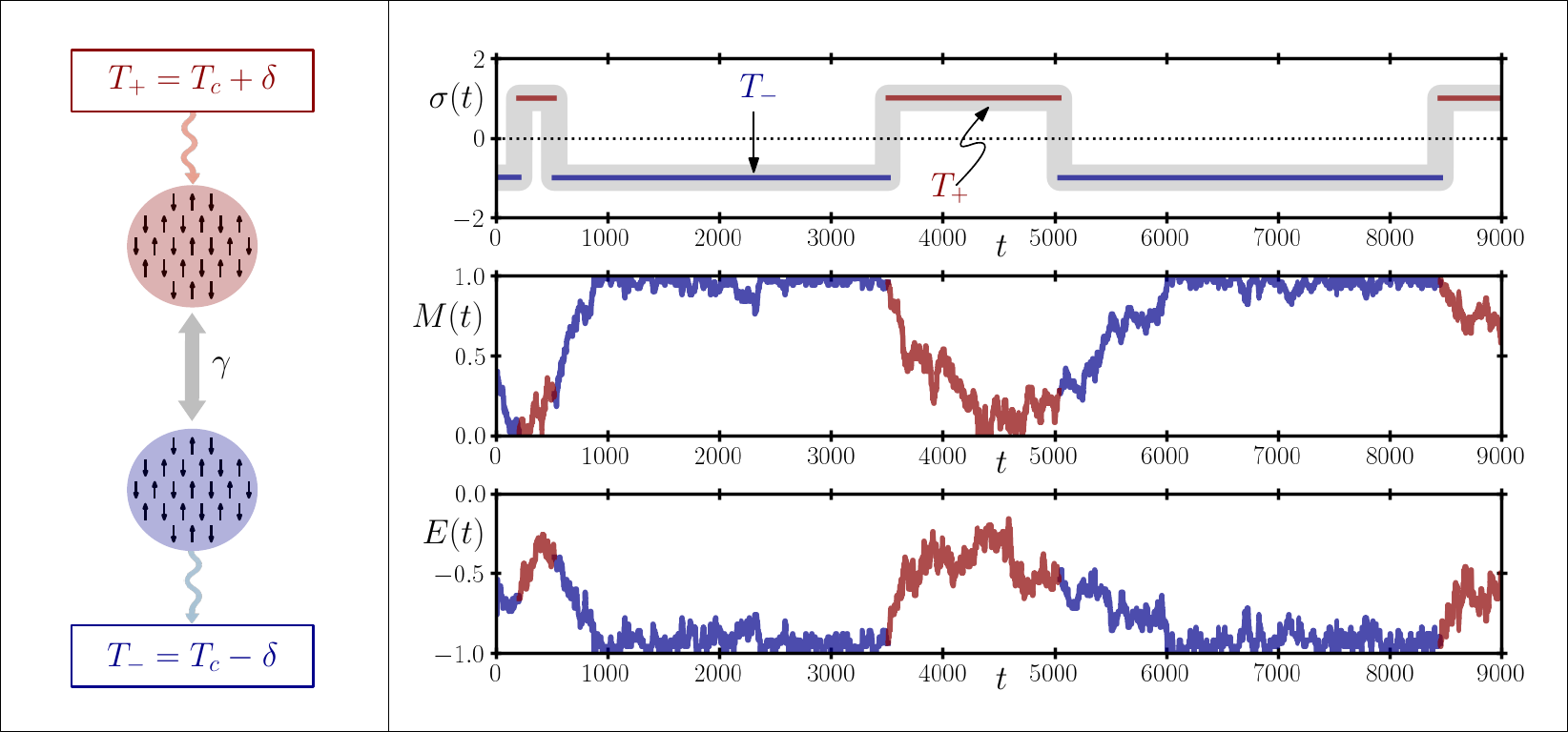}
    \caption{Schematic diagram of two-dimensional Ising magnet driven by stochastic two-temperature bath. The right panel shows typical trajectories of $\sigma(t)$, $M(t)$ and $E(t)$.}
    \label{fig:schematic}
\end{figure}

\section{Model and Results}\label{sec:model}
We consider a periodic square lattice of size $N=L \times L$ where each site $i$ is associated with a classical spin $s_i= \pm 1$. The spins interact via the nearest neighbour Ising Hamiltonian, where the energy of a configuration $C=\{s_1, s_2, \cdots s_N \}$ is given by,
\bea 
H(C) = - \sum_{\la ij \ra} s_i s_j. \label{eq:ham}
\eea 
The system is alternately coupled to two thermal reservoirs at temperatures $T_c - \delta$ and $T_c + \delta$, and the switching between them is governed by a dichotomous stochastic process with constant rate $\gamma$ [see Fig.~\ref{fig:schematic}]. The configuration of the system evolves via a Glauber-like local detail balanced single spin-flip dynamics where a randomly chosen spin flips with a rate, 
\begin{align}
     w_{\sigma}(\Delta E) = \frac{\psi_{0}}{2}\Big[(1+\sigma)e^{-\beta_{+}\Delta E/2}+(1-\sigma)e^{-\beta_{-}\Delta E/2}\Big],\label{eq:rate}
\end{align}
where $\Delta E$ is the change in the energy due to the proposed spin flip and $\beta_\sigma \equiv 1/T_{\sigma} = 1/(T_c + \sigma \delta)$ denotes the inverse temperature with $\sigma= \pm 1$. The intermittent change in the temperature is incorporated via the dichotomous process $ \sigma \to - \sigma$ with rate $\gamma$.
Note that, $\psi_0$ is taken to be a constant, independent of temperature, and characterises the inverse of the typical time-scale of spin flips in the system. Thus for a given $\sigma$ the system evolves via ordinary Glauber dynamics at a temperature $T_\sigma$.

For $\delta=0$ the system is equivalent to the ordinary Ising model at a constant temperature $T_c$. In that case, for any finite size $L$, the system reaches an equilibrium state with average magnetization $M_c$ and average energy $E_c$, that are independent of $\gamma$. In the thermodynamic limit, $L\to\infty$, the aforementioned equilibrium state becomes critical. For $\delta>0$, the system is expected to reach a nonequilibrium stationary state (NESS) the properties of which depends explicitly on $\gamma$ and $\delta$. Clearly, this NESS remains invariant under the transformation $\delta\to-\delta$. Our goal in this work is to characterize this NESS. In particular, we focus on the behaviour of magnetization $M$ per site and energy $E$ per bond, defined as,
\begin{align}
    M=\frac{1}{N}\left|\sum_{i=1}^{N}s_{i}\right|,\quad E = \frac{1}{2 N}\sum_{\la ij\ra}s_{i}s_{j}.
\end{align}
Typical time evolutions of $M(t)$ and $E(t)$ are shown in Fig.~\ref{fig:schematic}. Using numerical simulations, we measure the mean and variance of these observables in the NESS, and find that these quantities exhibit certain unusual features, including non-monotonic behaviour as a function of $\gamma$ for a finite sized system. We analytically characterise the behaviour in the limiting scenarios of small and large temperature switching rate $\gamma$. Before going into the details, we provide a brief summary of our main results.

\begin{figure}[t]
\centering
\includegraphics[width=7.5cm]{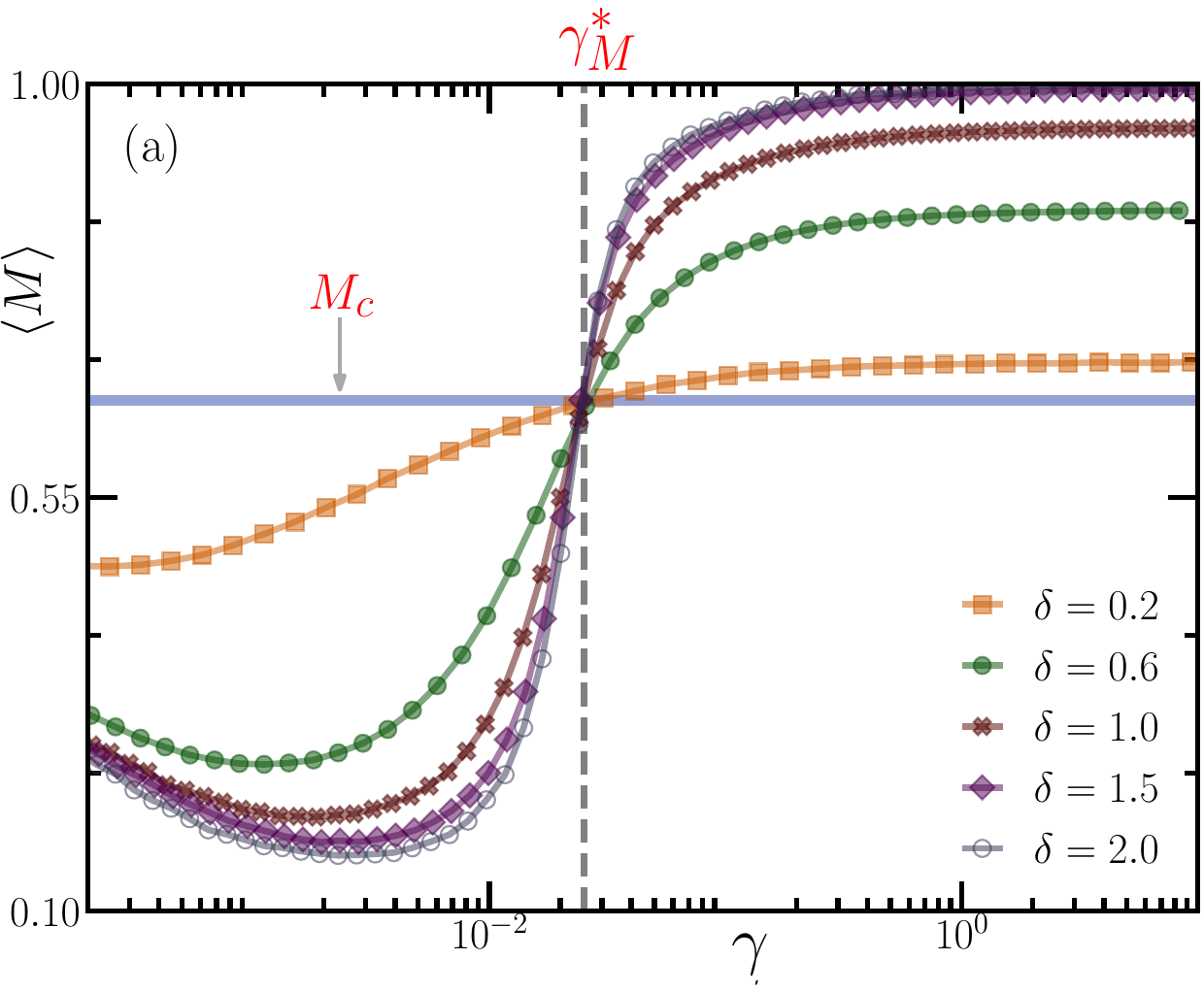}\hspace{0.cm}\includegraphics[width=7.5cm]{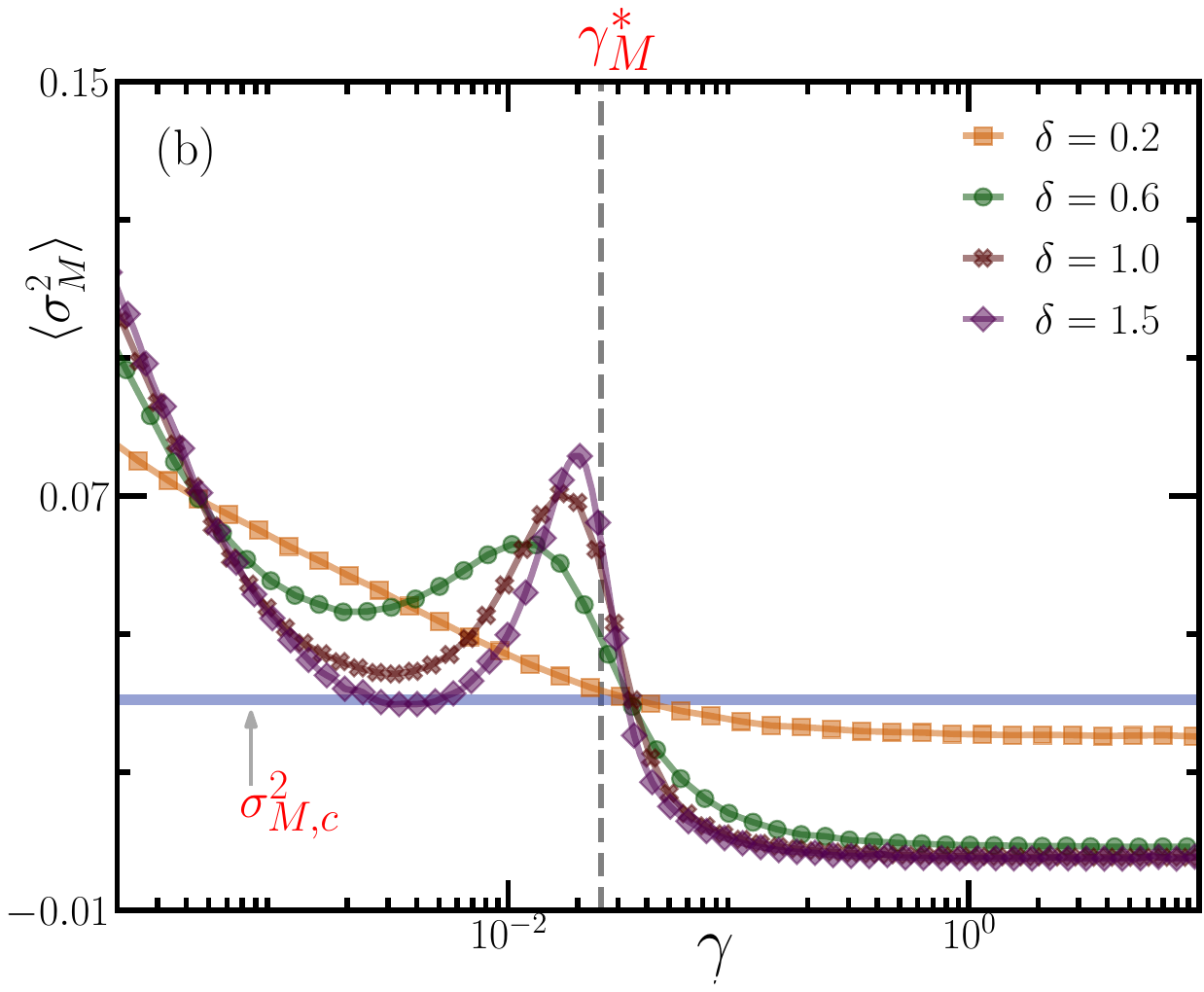}
    \caption{Plots of (a) average and (b) variance of the magnetization as functions of flip-rate $\gamma$ for different values of $\delta$. The horizontal solid lines indicate the corresponding equilibrium values at $T=T_{c}$ and the vertical dashed lines indicate the $\gamma^{\star}_{M}$. Here we have used $L=32$ and  $\psi_{0}=0.01$.} 
    \label{fig:Mg}
\end{figure}

\begin{itemize}

    \item  For large $\delta \lesssim T_c $, both average magnetization $\la M\ra$ and average energy $\la E \ra$ show non-monotonic behaviour with $\gamma$ --- while $\la M\ra_\gamma$ first decreases and then increases,  $\la E(\gamma)\ra$ first shows an increase, eventually decreasing as $\gamma$ is increased [see Figs.~\ref{fig:Mg}(a) and \ref{fig:Eg}(a)]. For small $\delta$, the non-monotonicity disappears, where $\la M \ra_\gamma$ $\big(\la E \ra_\gamma\big)$ monotonically increases (decreases) as a function of $\gamma$. Moreover, the variance of magnetization $\sigma_{M}^2$ shows a highly non-trivial behaviour with $\gamma$, whereas the variance of energy $\sigma_{E}^2$ decreases monotonically with $\gamma$ [see Figs.~\ref{fig:Mg}(b) and \ref{fig:Eg}(b)].
    
    \item In the small $\gamma$ regime the typical duration between two successive temperature switching events is much larger compared to the typical relaxation time of the Ising model at a given temperature. Using this time-scale separation, we obtain the behaviour of $\la M \ra_\gamma$ and $\la E \ra_\gamma$ in the small $\gamma$ regime. This also explains the emergence of the non-monotonicity for large $\delta\lesssim T_c$ [see Fig.~\ref{fig:O_msg_esg}].

    \item Investigating the behaviour of $\la M \ra$ and $\la E \ra$ as a functions of $\delta$, we find that for small $\delta$, 
    \begin{align}
        \la M \ra_\delta-M_{c}= \chi_{2}^{M}(\gamma)~\delta^{2}+O(\delta^{4}),\quad\text{and}\quad\la E\ra_\delta-E_{c} = \chi_{2}^{E}(\gamma)~\delta^{2}+O(\delta^{4}).
    \end{align}
    where $M_{c}$ and $E_{c}$ denote the equilibrium values of average magnetization and energy, respectively, at $T=T_{c}$. Using dynamical response theory, we investigate the $\gamma$ dependence of the second order susceptibilities $\chi_{2}^{M}$ and $\chi_{2}^{E}$ [see Fig.~\ref{fig:Og_smdelT}]. Interestingly, we find that both $\chi_{2}^{M}$ and $\chi_{2}^{E}$ change sign as $\gamma$ is increased beyond certain threshold values. 

    \item We show that, in the $\gamma \to \infty$ limit, the spin-flip rate $w(\Delta E)$ can be expressed as that of an equilibrium Ising model with an effective temperature 
    \bea 
        T_\text{eff} = T_{c}\Big(1 - \frac{\delta^{2}}{T_{c}^2}\Big).
    \eea 
    In fact, we find that for large $\gamma \gg \psi_0$, an effective equilibrium picture emerges, with a $\gamma$ and $\delta$ dependent effective temperature [see Fig.~\ref{fig:Tg}]. However, the inherent nonequilibrium nature of the process is apparent from a non zero energy current flowing through the system [see Fig.~\ref{fig:jg_jdelT}].
    
\end{itemize}

In the next section, we study the behavior of the NESS in detail.

\section{Characterization of the NESS}\label{sec:NESS}
To characterise the stationary state of this stochastic two-temperature nonequilibrium Ising model, we first focus on the behaviour of mean and variance of magnetisation and energy. For such a nonequilibrium setting it is generally hard to compute these quantities analytically. Hence we take recourse to Monte-Carlo simulations to explore the behaviour of these quantities for a finite system in the $\delta-\gamma$~plane. For the sake of completeness, we briefly describe the details of the simulation. 

We initialize the system in a random configuration, where each spin $s_{i}$ is chosen to be $\pm 1$ with equal probability $1/2$. Additionally, we initialize $\sigma=\pm 1$ with equal probability $1/2$, which corresponds to a temperature $T_{\sigma}=T_c+\sigma\delta$. The temperature remains constant for a time $\tau$ which is drawn from an exponential distribution $P(\tau)=\gamma e^{-\gamma \tau}$, during which the spins evolve via Glauber dynamics---a randomly chosen spin flips with the rate $w(\Delta E)$ defined in \eref{eq:rate}. After time $\tau$ is elapsed, $\sigma\to-\sigma$ resulting in a temperature switch. The temperature now remains constant at $T_{-\sigma}$ for a time drawn from $P(\tau)$. This dynamics eventually leads to an NESS where we measure the relevant observables.

Let us first investigate the behaviour of magnetization and energy as functions of $\gamma$. Figure~\ref{fig:Mg}(a) shows a plot of $\la M \ra$ versus $\gamma$ for different values of $\delta$. Evidently, for $\delta=0$, the magnetization remains constant at $\la M \ra=M_{c}$, where $M_{c}$ is the average value of magnetization for the equilibrium $L\times L$ Ising model at temperature $T_{c}$. For $\delta>0$, average magnetization shows a non-trivial dependence on $\gamma$---for small values of $\gamma$, $\la M \ra<M_{c}$, while for large $\gamma$, $\la M\ra>M_{c}$. In fact, it appears that for some intermediate $\gamma=\gamma^{\star}_{M}$, $\la M \ra=M_{c}$ irrespective of $\delta$, which corresponds to the crossing point of the different curves in Fig.~\ref{fig:Mg}(a). Remarkably, for large values of $\delta$, i.e., as the system gets further away from equilibrium, the dependence of $\la M \ra$ on $\gamma$ becomes non-monotonic in the $\gamma<\gamma^{\star}_{M}$ regime. However, for $\gamma>\gamma^{\star}_{M}$, the magnetization monotonically increases for all values of $\delta$ indicating an overall increase in the spin alignment. Unusual signatures are also exhibited by the variance of the magnetization, $\sigma_M^2=\la M^{2}\ra-\la M\ra^{2}$, which shows a peak near $\gamma=\gamma^{\star}_{M}$ that becomes sharper with increasing $\delta$ [see Fig.~\ref{fig:Mg}(b)]. 

\begin{figure}[t]
    \centering
    \includegraphics[width=8.5cm]{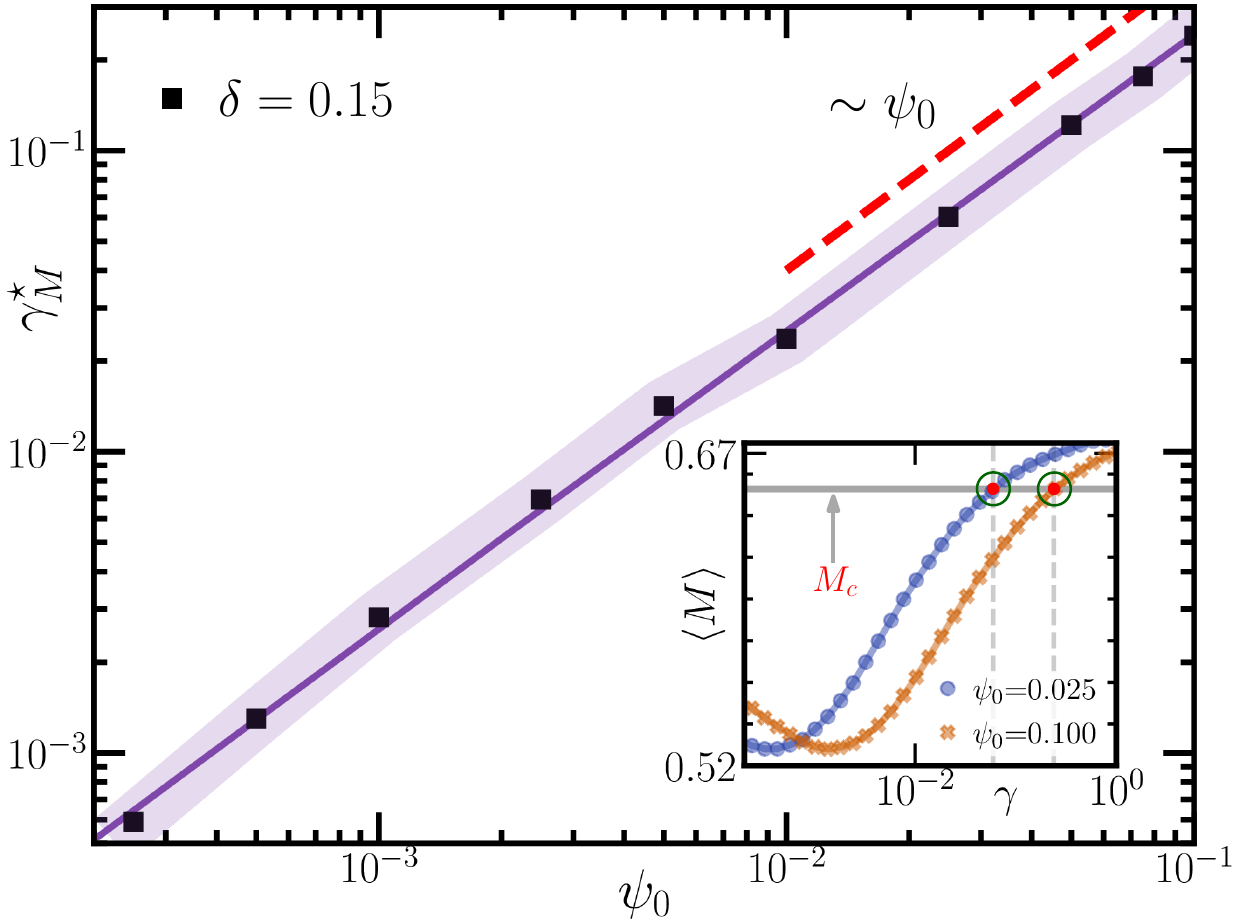}
    \caption{Plot of $\gamma^{\star}_{M}$ versus $\psi_{0}$ for a fixed small $\delta$. The symbols indicate data obtained from numerical simulation and the solid line denotes the best fit straight line. The shaded region indicates the error margin corresponding to the symbols. The inset illustrates how $\gamma^{\star}_{M}$ is extracted from the $\la M\ra$ vs $\gamma$ curves. Here we have used $L=32$ and $\delta=0.15$.} 
    \label{fig:gstar_psi}
\end{figure}

Thus from Fig.~\ref{fig:Mg} it is apparent that the stochastic temperature switching leads to some intriguing features in the behaviour of the magnetization. These features are most prominent when the switching rate is smaller than the threshold value $\gamma^{\star}_{M}$. We also investigate how this threshold changes with the typical  intrinsic spin-flip time-scale. Figure~\ref{fig:gstar_psi} shows a plot of $\gamma^{\star}_{M}$ as a function of $\psi_{0}$ which shows a linear increase. This indicates that the effect of stochastic evolution of the temperature is most striking when the typical time-scale of temperature switching is smaller than that of the spin flip.  

\begin{figure}[t]
\centering
\includegraphics[width=7.5cm]{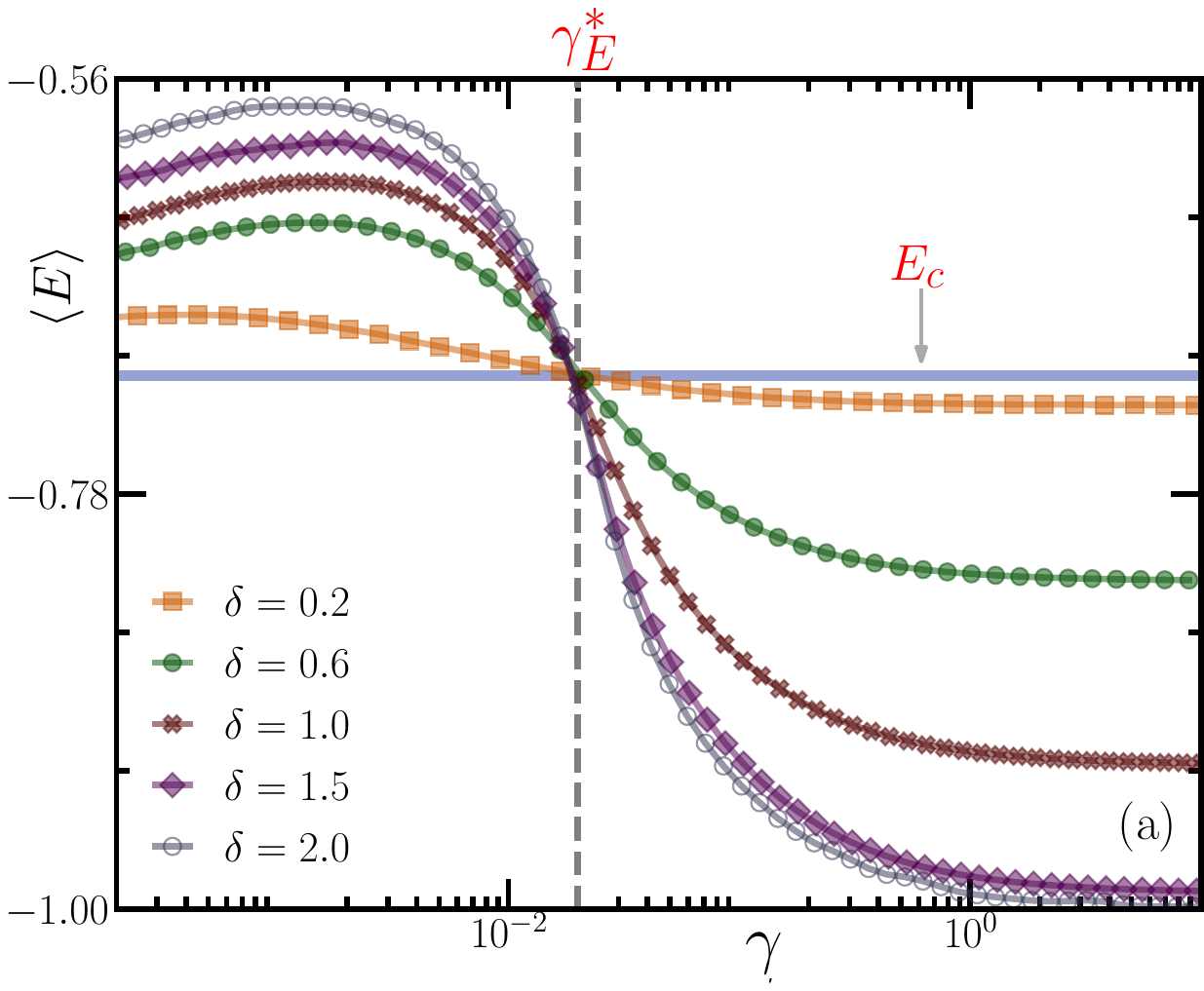}\includegraphics[width=7.5cm]{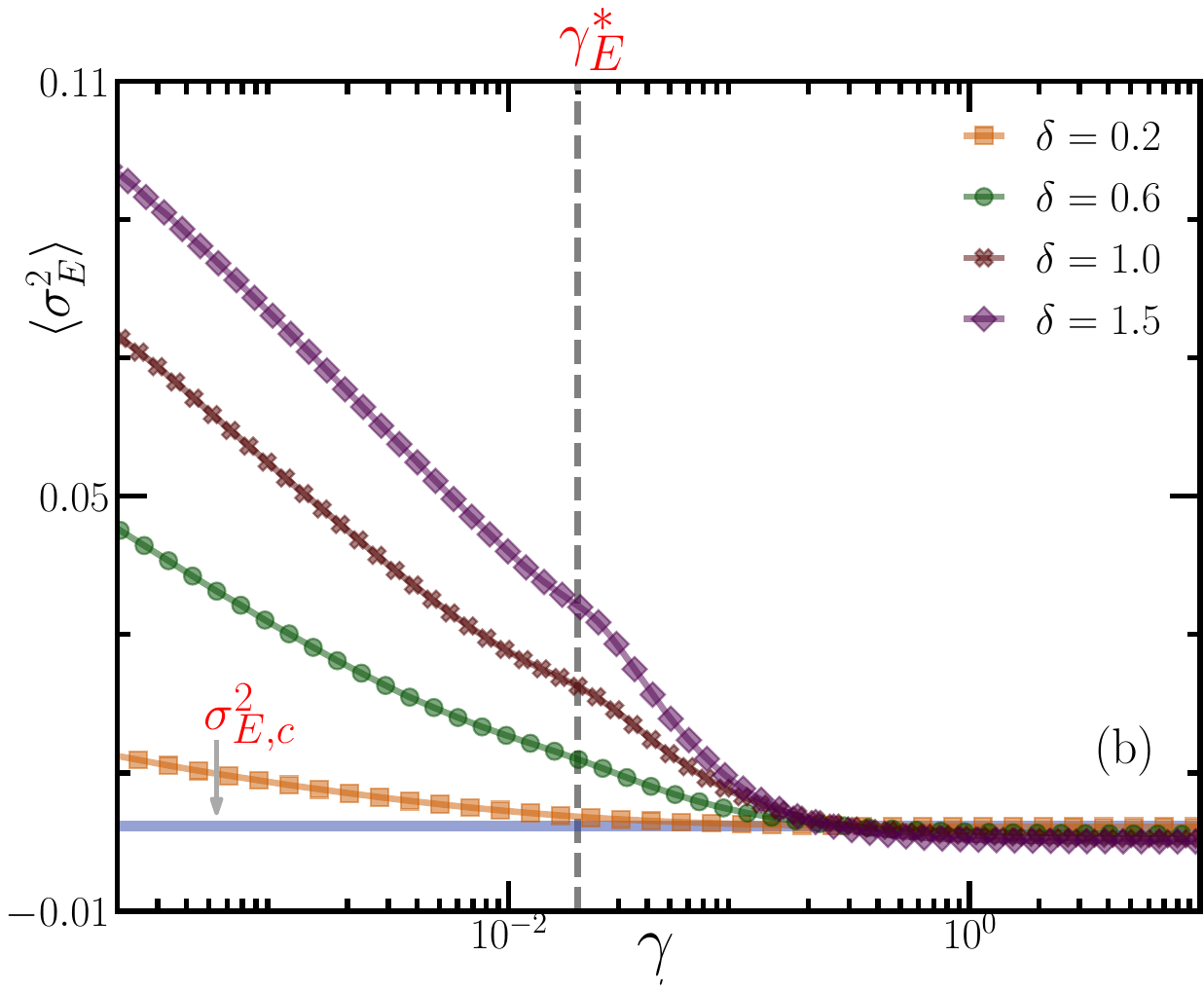}
    \caption{Plots of (a) average and (b) variance of the energy as functions of flip-rate $\gamma$ for different values of $\delta$. The horizontal solid lines indicate the corresponding equilibrium values at $T=T_{c}$ and the vertical dashed lines indicate the $\gamma^{\star}_{E}$. Here we have used $L=32$ and  $\psi_{0}=0.01$.}  
    \label{fig:Eg}
\end{figure}

We also explore the behaviour of mean and variance of the energy per bond as functions of $\gamma$. Figure~\ref{fig:Eg}(a) shows plots of $\la E \ra$ versus $\gamma$---similar to magnetization, the curves corresponding to different values of $\delta$ show a crossing a threshold value $\gamma^{\star}_{E}$. Note that, $\gamma^{\star}_{E}$ is different from $\gamma^{\star}_{M}$ although they are of the same order. Contrary to the magnetization, the average energy $\la E \ra>E_{c}$, the value of average energy at $\delta=0$, for $\gamma<\gamma^{\star}_{E}$, while $\la E \ra<E_{c}$ for $\gamma>\gamma^{\star}_{E}$. It is clear from Fig.~\ref{fig:Eg}(a) that $\la E \ra$ also shows a nonmonotonicity for large values of $\delta$ in the $\gamma<\gamma^{\star}_{E}$ regime. On the other hand, the variance of energy $\sigma^2_{E}=\la E^{2} \ra - \la E \ra^{2}$ decreases monotonically with $\gamma$ [see Fig.~\ref{fig:Eg}(b)]. 

\begin{figure}
    \centering
    \includegraphics[width=7.5cm]{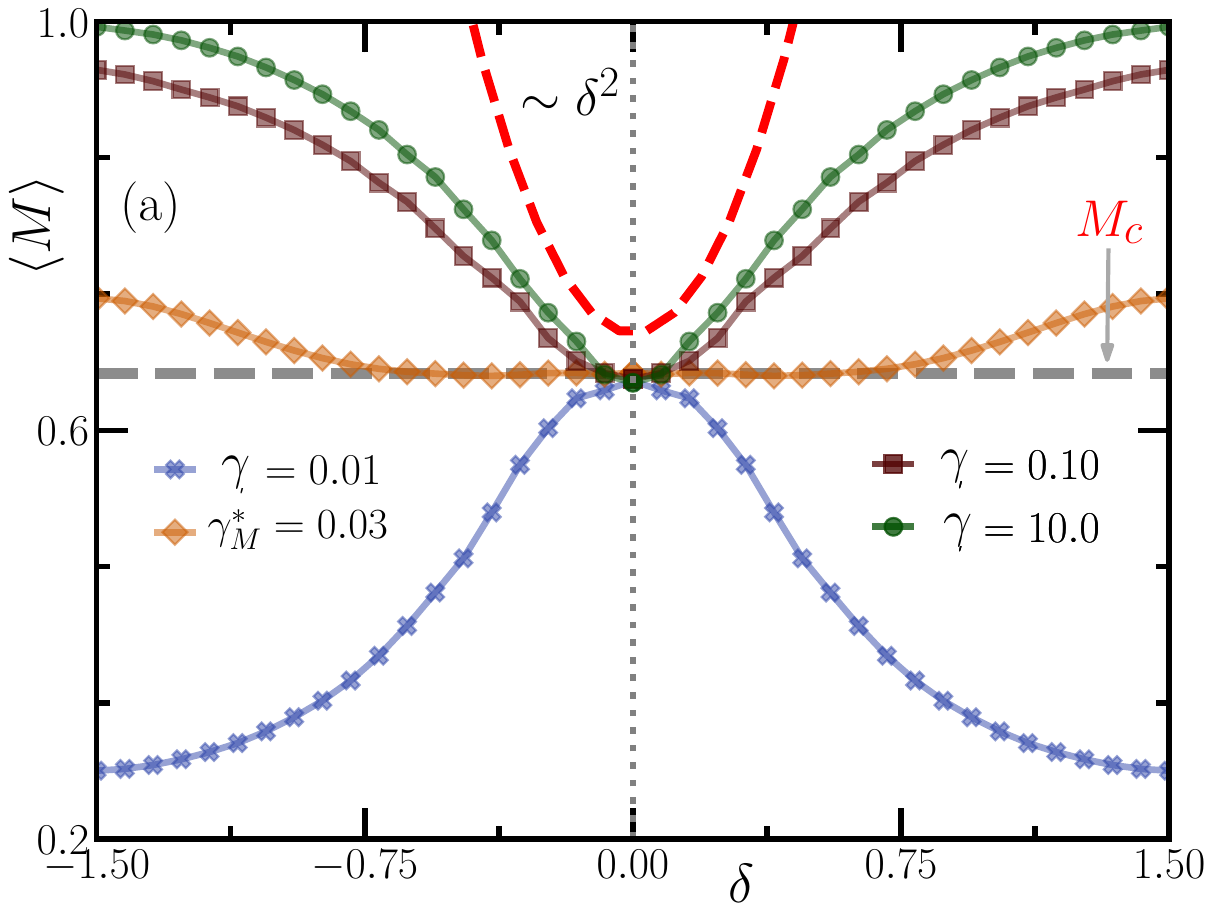}\hspace{-0.1cm}\includegraphics[width=7.5cm]{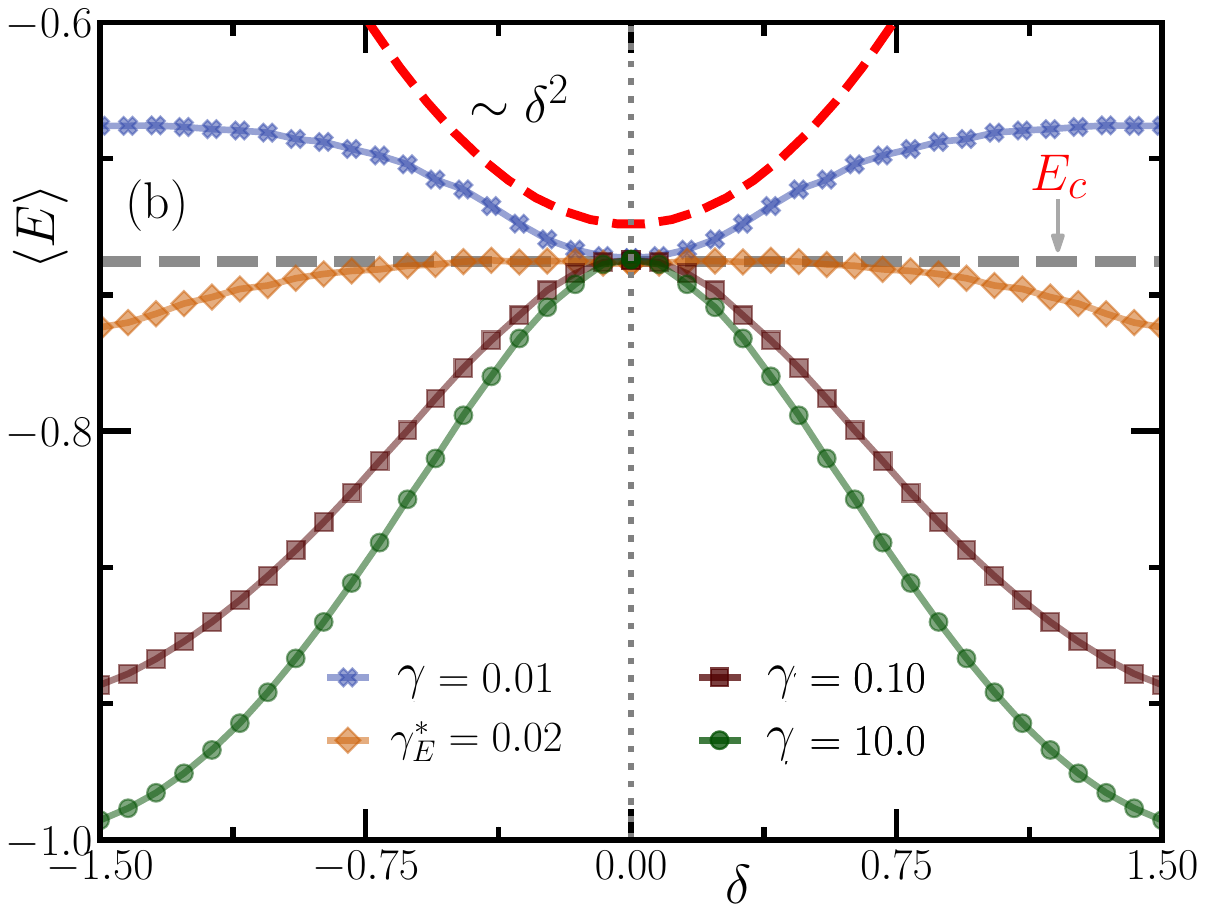}
    \caption{Plots of (a) average magnetization $\langle M \rangle$ and (b) average energy $\langle E \rangle$ as functions of $\delta$ for different flip rates $\gamma$. The horizontal dashed lines mark the corresponding equilibrium values $M_{c}$ and $E_{c}$ at $T = T_{c}$. Red dashed parabolas are shown as guides to the eye, highlighting the quadratic behaviour at small $\delta$. Here we have used $L=32$ and $\psi_0=0.01$.}
    \label{fig:OdelT}
\end{figure}

We also investigate how average magnetization and energy change with $\delta$. As shown in Fig.~\ref{fig:OdelT} both $\la M \ra$ and $\la E \ra$ are symmetric functions of $\delta$. The change in the behaviour of the system across the threshold values of $\gamma$ is manifest in the change of curvature of the $\la M \ra$ and $\la E \ra$ curves. The average magnetization $\la M \ra$ is concave around $\delta=0$ when $\gamma<\gamma^{\star}_{M}$ and becomes convex for $\gamma>\gamma^{\star}_{M}$. In contrast, $\la E \ra$ is convex for $\gamma<\gamma^{\star}_{E}$ and turns concave for $\gamma>\gamma^{\star}_{E}$.

In the following, we first focus on the two limiting scenarios, $\gamma\ll\psi_{0}$ and $\delta\ll T_{c}$, to understand the physical origin of the non-trivial $\gamma$ and $\delta$ dependence of magnetization and energy. Finally we focus on the effective temperature picture emerging in the regime $\gamma \gg \psi_{0}$.

\subsection{\texorpdfstring{Small $\gamma$ Regime}{Small gamma Regime}}\label{subsec:SG}
To understand the behaviour of the NESS in the small $\gamma$ regime, it is useful to consider a renewal approach~\cite{coxx_renewal,ross1995stochastic}. Since the time evolution of the system between two consecutive temperature switches follows ordinary Ising dynamics at a constant temperature, we can write a last renewal equation, 
\begin{align}
    P_{\gamma}(C,t|C_{0}) &= e^{-\gamma t}\sum_{\sigma_{0}}\rho(\sigma_{0})\mathbb{P}_{T_{\sigma_{0}}}(C,t|C_{0})+\gamma\sum_{\sigma,C'}\rho(\sigma)\intop_{0}^{t}dt'~e^{-\gamma t'}\mathbb{P}_{T_{\sigma}}(C,t'|C^{\prime})P_{\gamma}(C^{\prime},t-t'|C_{0}), \label{eq:P_ren}
\end{align}
where, $P_{\gamma}(C,t|C_{0})$ denotes the probability that the system is in spin configuration $C$ at time $t$ starting from $C_{0}$. This equation can be understood by considering contributions from trajectories with no temperature switch, indicated by the first term and atleast one temperature switch, indicated by the second term. In the second term, $t'$ denotes the time elapsed since the last temperature switch to the value $T_{\sigma}$, $\mathbb{P}_{T}(C,t'|C^{\prime})$ denotes the probability that ordinary Ising model at constant temperature $T$ is in configuration $C$ at time $t'$ starting from $C^{\prime}$. Moreover, $P_{\gamma}(C^{\prime},t-t'|C_{0})$ denotes the probability that the system has configuration $C^{\prime}$ at the time of last temperature switch, $t-t'$, starting from $C_{0}$.
Here we have assumed that at $t=0$, the system can be at temperature $T_{\pm}$ with equal probability $1/2$, i.e., $\sigma_{0}=\pm1$ with $\rho(\sigma_{0}) = 1/2$.

\begin{figure}[t]
    \centering
    \includegraphics[width=7.5cm]{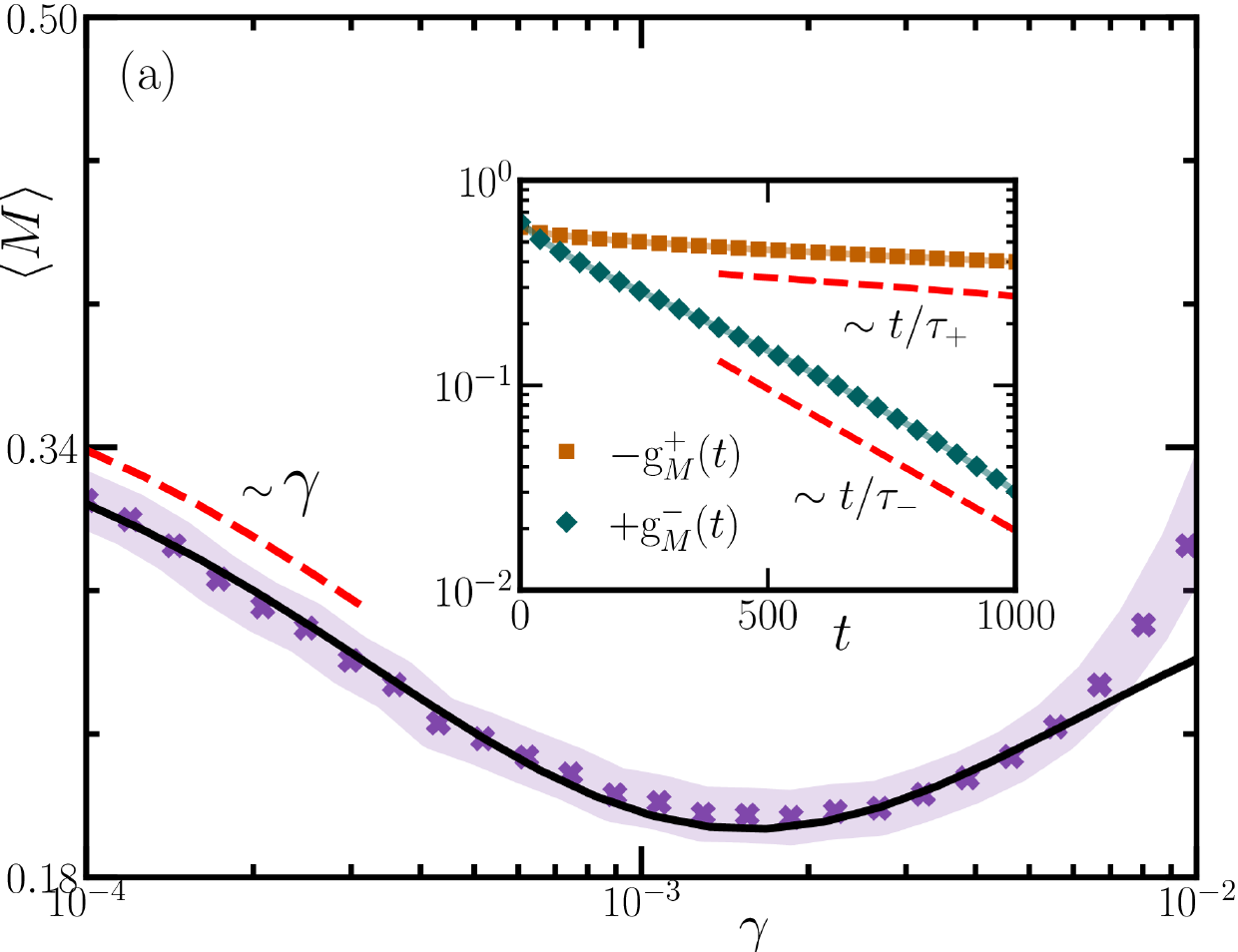}\includegraphics[width=7.5cm]{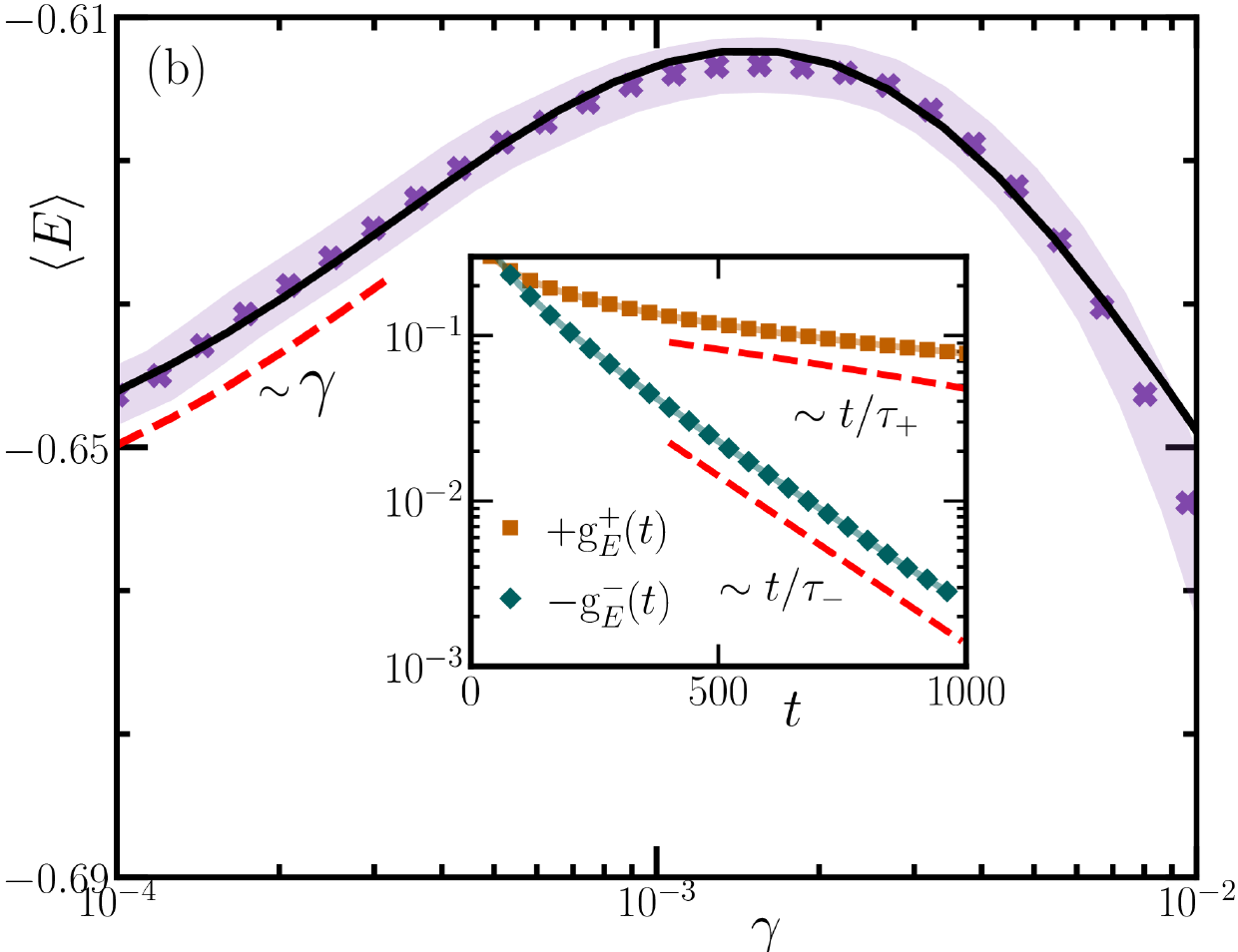}
    \caption{Small $\gamma$ regime: (a) Average magnetisation $\la M \ra$ and (b) average energy $\la E \ra$ as functions of $\gamma$ in the regime $\gamma\ll\psi_{0}$. The symbols indicate the data obtained from numerical simulations with the shaded region indicating the error margin. The solid lines correspond to \eref{eq:sg_O} with the parameters $b_{\pm}$ and $\tau_{\pm}$ extracted by fitting $g_{O}^{\sigma}(t)$ to the empirical form \eref{eq:relax_approx} [see the insets]. Here we have used $L=32, \psi_0=0.01$ and $\delta=1.0$.}
    \label{fig:O_msg_esg}
\end{figure}

In general, it is hard to solve the integral \eref{eq:P_ren} to obtain $P_\gamma(C,t)$ explicitly. However, in the small $\gamma$ limit, this equation can be used to obtain the stationary value of the average of any observable $\la O \ra$ as discussed below. For ordinary 2D Ising model with Glauber dynamics, the relaxation time $\tau_{r}\sim\psi_{0}^{-1}L^{2}$. For small values of $\gamma\ll\tau_{r}^{-1}$, the typical duration between consecutive temperature flips $\gamma^{-1}$ is much larger compared to the typical relaxation time of ordinary Ising model. Consequently, the system approaches close to equilibrium at temperatures $T_{+}$ and $T_{-}$ between successive temperature switches. In this regime, $\mathbb{P}_{T_{\sigma}}(C,t'|C^{\prime})$ becomes independent of the initial spin configuration $C^{\prime}$ and \eref{eq:P_ren} reduces to,
\begin{align}
    P_{\gamma}(C,t) \simeq \frac{1}{2}e^{-\gamma t}\Big[\mathbb{P}_{T_{+}}(C,t)+\mathbb{P}_{T_{+}}(C,t)\Big]+\frac{\gamma}{2}\intop_{0}^{t}dt'~e^{-\gamma t'}\Big[\mathbb{P}_{T_+}(C,t')+\mathbb{P}_{T_-}(C,t')\Big], \label{eq:P_ren_simp}
\end{align}
The expectation values of any single-time observable $O(C)$, like magnetization and energy, in the stationary state, can thus be expressed as,
\begin{align}
    \la O\ra_{\gamma} &\equiv\lim_{t\to\infty}\sum_{C} O(C) P_{\gamma}(C,t)
    = \frac{\gamma}{2}\intop_{0}^{\infty}dt'~e^{-\gamma t'}\Big[ \la O(t') \ra_{\rm eq}^{+} + \la O(t') \ra_{\rm eq}^{-}\Big].\label{eq:O_st}
\end{align}
The above expression allows us to compute the stationary average $\la O \ra_\gamma$ of any observable using its equilibrium relaxation behaviour $\la O(t)\ra_{\rm eq}^{\pm}$ for ordinary Ising model at temperature $T_{\pm}$. For any finite-sized system at temperature $T_{\sigma}$, the observables typically show an exponential relaxation to a equilibrium value $\la O\ra_{\rm eq}^{\sigma}$. Thus we can assume, 
\begin{align}
    \la O(t) \ra_{\rm eq}^{\sigma}=\la O\ra_{\rm eq}^{\sigma}+b_{\sigma} e^{-t/\tau_{\sigma}}, \label{eq:relax_approx}
\end{align}
where $\tau_{\sigma}$ denotes the relaxation time-scale at temperature $T_{\sigma}$, and $b_{\sigma}$ is an empirical constant. Substituting \eref{eq:relax_approx} in \eref{eq:O_st} and performing the integration over $t'$, we get,
\begin{align}
    \la O\ra_\gamma &= \frac{1}{2}\Big[\la O \ra_{\rm eq}^{+} + \la O \ra_{\rm eq}^{-}\Big] + \frac{\gamma}{2}\Big[\frac{b_{+}\tau_{+}}{1+\gamma \tau_{+}}+\frac{b_{-}\tau_{-}}{1+\gamma\tau_{-}}\Big].\label{eq:sg_O}
\end{align}
Clearly, in the limit $\gamma\to0$, we have $\la O \ra_{\gamma}=\Big[\la O \ra_{\rm eq}^{+} + \la O \ra_{\rm eq}^{-}\Big] / 2$, as expected. Equation \eqref{eq:sg_O} allows us to determine the small $\gamma$ behaviour of $\la M \ra$ and $\la E \ra$ in the NESS, using the corresponding $\la O \ra_{\rm eq}^{\sigma}$, $b_{\sigma}$ and $\tau_\sigma$ values extracted from the equilibrium relaxation dynamics at $T_{\sigma}$. This can be done by looking at,
\bea 
g_{O}^{\sigma}(t)=\la O(t)\ra_{\rm eq}^{\sigma} - \la O\ra_{\rm eq}^{\sigma}
\eea
for ordinary Ising model at $T_\sigma$ and extracting the parameters $\tau_\sigma$ and $b_\sigma$, which depend on the observable, by fitting it to the form given in \eref{eq:relax_approx}. We estimate these parameters for a given $\delta$, for both energy and magnetization, from their equilibrium relaxation dynamics and compute their $\gamma$-dependence. Figure ~\ref{fig:O_msg_esg} compares these predicted behaviours with $\la M \ra_{\gamma}$ and $\la E \ra_\gamma$ obtained from numerical simulations, which show an excellent match. This validates our prediction \eref{eq:sg_O} and the assumptions leading to it.

This analysis also explains the non-monotonic behaviour of $\la M \ra_{\gamma}$ and $\la E \ra_{\gamma}$. In the small $\gamma$ regime, at each temperature switching event $\sigma\to-\sigma$, the expectation value $\la O(t) \ra_{\gamma}$ starts from $\la O\ra_{\rm eq}^{\sigma}$, eventually reaching $\la O\ra_{\rm eq}^{-\sigma}$ before the next switch [see Fig.~\ref{fig:schematic}]. Thus, $b_{+}$ and $b_{-}$ must have opposite signs. For example, for magnetisation $b_{+}>0$ and $b_{-}<0$, while for energy $b_{+}<0$ and $b_{-}>0$. It is then easy to show that, for $\tau_{+}\neq\tau_{-}$, \eref{eq:sg_O} has an extremum at some intermediate value of $\gamma$. Since, the relaxation in the ferromagnetic phase is much slower compared to that in the paramagntic phase, $\tau_- >\tau_{+}$, with their difference becoming larger as $\delta$ is increased. This difference in the relaxation time-scales leads to the non-monotonic variation of $\la M \ra_{\gamma}$ and $\la E \ra_{\gamma}$ in the small $\gamma$ regime, as shown in Fig.~\ref{fig:O_msg_esg}.

\subsection{\texorpdfstring{Small $\delta$ regime}{Small delta regime}}\label{subsec:SD}
In this section, we study the behaviour of average energy and average magnetization for small values of $\delta$, i.e., when $\delta/T_{c}\ll 1$. The $\delta= 0$ limit corresponds to the case of the equilibrium Ising model at temperature $T_{c}$. Therefore, the small-$\delta$ behaviour of the system can be explored by applying the dynamical response theory~\cite {PhysRevLett.96.240601,frenetic_aspect_response,frenesy_maes,maes2020response} around the equilibrium limit, treating $\delta$ as the perturbation parameter. Since for $\delta=0$, the temperature flip dynamics is decoupled from the spin dynamics, the unperturbed system corresponds to two independent equilibrium processes, namely, the Ising dynamics at temperature $T_c$ and a dichotomous process, $\sigma\to-\sigma$ with rate $\gamma$. For non-zero values of $\delta$ the coupled dynamics reaches an NESS which remains invariant under the transformation $\delta\to-\delta$ and the expectation value of any observable $\la O \ra_{\delta}$ must be an even function of $\delta$ [see Fig.~\ref{fig:OdelT}]. Therefore, for small $\delta$, we can write,
\begin{align}
    \la O \ra_{\delta}=\la O \ra_{\rm eq}+\delta^2\chi_2+\mathscr{O}(\delta^4),
    \label{eq:master_response}
\end{align}
where $\la O \ra_{\rm eq}$ denotes the equilibrium value of the observable and  $\chi_{2}$ represents the second order response coefficient.  In the following, we compute this coefficient using the non-linear dynamical response theory, where frenetic contributions play a significant role~\cite{frenetic_aspect_response, frenesy_maes}.  
For a generic observable $O(\omega)$, which depends on the trajectory $\omega$ of the system over a time-interval $[0,t]$, the second order response coefficient is given by, 
\begin{align}
    \chi_{2}(\omega)=\frac{1}{2}\left\la O(\omega)\left[-D''(\omega)+\frac{1}{2}S''+(D')^2(\omega)+\frac{1}{4}(S')^2(\omega)-D'(\omega)\,S'(\omega)\right] \right\ra_{\rm eq},\label{eq:chi2_def}
\end{align}
where  $S'(\omega)$ and $D'(\omega)$ denote the first-order excesses in entropy and frenesy, generated due to the perturbation. The corresponding second-order contributions are indicated by $S''(\omega)$ and $D''(\omega)$. For Markov jump-processes, $S(\omega)$ and $D(\omega)$ along the trajectory $\omega$, have the general form~\cite{maes2020response},
\begin{align}
    S(\omega) &= \sum_{\ell}\ln{\frac{k(x_{t'_\ell},x_{t'_{\ell+1}})}{k(x_{t'_{{\ell+1}}},x_{t'_\ell})}}\label{eq:entropy}\\
    D(\omega) &= \intop_{0}^{t}dt'~\big[\lambda(x_{t'})-\lambda_{0}(x_{t'})\big]-\frac{1}{2}\sum_{\ell}\ln{\Big[k(x_{t'_\ell},x_{t'_{\ell+1}})k(x_{t'_{\ell+1}},x_{t'_\ell})\Big]}\label{eq:frenesy}
\end{align}
where $k(x_{t'_\ell},x_{t'_{\ell+1}})$ is the transition rate for the $\ell$-th jump from the configuration $x_{t'_\ell}$ to $x_{t'_{\ell+1}}$, $\lambda(x)=\sum_{y\neq x}k(x,y)$ denotes the escape rate from the state $x$ and $\sum_{\ell}$ denotes a sum over all transitions $\ell$ along the trajectory $\omega$. 

Since we are interested in the response coefficient of single-time observable like $M(t)$ and $E(t)$, it is useful to consider a time-antisymmetric observable of the form $O(\omega,t)=O(t)-O(0)$ in \eref{eq:chi2_def}. In this case, the the correlations $\la (D')^{2}O\ra_{\rm eq}$, $\la (S')^2O\ra_{\rm eq}$, and $\la D''O\ra_{\rm eq}$
vanish due to the time-reversal invariant nature of equilibrium dynamics. Consequently, to second order in $\delta$, we have,
\begin{align}
    \la O(t) \ra_{\delta} - \la O \ra_{\rm eq} = \delta^{2}\Big[\frac{1}{2}\la S^{\prime\prime} O(t) \ra_{\rm eq}- \la S^{\prime} D^{\prime} O(t) \ra_{\rm eq}\Big].\label{eq:response_chi2}
\end{align}
Clearly, the second order response coefficient is a sum of two terms---the first one, $\la S''O(t)\ra_{\rm eq}$ is purely entropic while the second one, $\la S'D'O(t)\ra_{\rm eq}$ involves the excess dynamical activity.

To proceed further, we note that for our two-temperature Ising model, the trajectory is characterised by $\omega=\{C(t'), \sigma(t'); 0<s<t\}$, where $C(t')$ denotes the spin configuration and $\sigma(t')$ indicates the instantaneous temperature $T(t')=T_c+\sigma(t')\delta$ at time $t'$. The transition rates are given by,
\begin{align}
        k\big(\{C,\sigma\}, \{C^{\prime}, \sigma\}\big) &=
            \psi_{0}~\exp\Big(-\frac{\beta_{\sigma}}{2}\big[H(C^{\prime}) - H(C)\big]\Big),~~&\text{for spin-flips,}\\
        k\big(\{C,\sigma\}, \{C, \sigma^{\prime}\}\big) &=\gamma ~~&\text{for temperature switches,}\label{eq:trans_rate}
\end{align}
where $H$ is the Hamiltonian of the Ising system given in \eref{eq:ham} and $\beta_{\sigma}=1/(T_c+\sigma \delta)$. Since we need to compute excesses of $S(\omega)$ and $D(\omega)$ with respect to $\delta$, it suffices to consider only the $\delta$ dependent terms,
\begin{align}
    S(\omega)&=-\sum_{\sigma}\beta_{\sigma}\Delta {E}^{\sigma}(t)+{\rm const.},\label{eq:closed_entropy}\\
    D(\omega)&=\psi_{0}\intop_{0}^{t}dt'~\sum_{k=1}^{L^{2}}e^{-\beta_{\sigma(t')}\Delta \varepsilon_{k}(t')/2}+{\rm const}.\label{eq:closed_frenesy}
\end{align}
Here, $\Delta E^{\sigma}(t)$ in \eref{eq:closed_entropy} is the total energy gained by the system from the reservoir at temperature $T_{\sigma}$ during the interval $[0, t]$. Moroever, $\Delta \varepsilon_{k}(t')$ in \eref{eq:closed_frenesy} denotes the energy cost associated with flip of the $k$-th spin in the configuration $\{C(t'),\sigma(t')\}$. 

The excess entropy can be computed by taking derivative of \eref{eq:closed_entropy} with respect to $\delta$ and then taking $\delta=0$. 
\begin{align}
    S^{\prime}(t) &=\frac{\mathscr{J}_{E}(t)}{T_{c}^{2}}=\frac{1}{T^{2}_{c}}\Big[\Delta E^{+}(t)-\Delta E^{-}(t)\Big],\label{eq:Sp_eq}\\
    S^{\prime\prime}(t)&=-\frac{U(t)}{T_{c}^{3}}=-\frac{2}{T_{c}^{3}}\Big[\Delta E^{+}(t)+\Delta E^{-}(t)\Big],\label{eq:Spp_eq}
\end{align}
where, $\mathscr{J}_{E}(t)$ denotes the time-integrated energy current flowing from the hot reservoir at temperature $T_{+}$ to the cold reservoir at temperature $T_{-}$. Moreover, $U(t)$ denotes the net energy gain of the system from both the reservoirs during the time interval $[0,t]$.

The excess frenesy can be similarly obtained from \eref{eq:closed_frenesy} and is given by,
\begin{align}
        D^{\prime}(t)&=\frac{\psi_0}{2T_{c}^2}\intop_{0}^{t}dt'~\sigma(t')\sum_{k=1}^{L^{2}}\Delta \varepsilon_{k}(t')\,e^{-\beta_{c}\Delta \varepsilon_{k}(t')/2}.\label{eq:Dp_eq}
\end{align}
Substituting the above equations in \eref{eq:response_chi2}, we explicitly obtain the second-order response coefficient, which can be conveniently expressed as,
\begin{align}
    \chi_2(t) &= \chi^{(1)}_{2}(t)+\chi^{(2)}_{2}(t),\label{eq:chi2_break}
\end{align}
where,
\begin{align}
    \chi_2^{(1)}(t) = -\frac{1}{T_{c}^{3}}\big\langle U(t) O(t) \big\rangle_{\rm eq},\label{eq:chi21}
\end{align}
denotes the purely entropic contribution, and 
\begin{align}
        \chi_2^{(2)}(t) = -\frac{\psi_{0}}{2T_c^4}\Big\langle O(t)\mathscr{J}_{E}(t)\intop_{0}^{t}dt'~\sigma(t')\sum_{k=1}^{L^{2}}\Delta \varepsilon_{k}(t')\,e^{-\beta_{c}\Delta \varepsilon_{k}(t')/2}\Big\rangle_{\rm eq}\label{eq:chi22}
\end{align}
involves the contribution from frenesy, explicitly depending on the dynamics. In the limit $t\to\infty$, $\chi_{2}(t)$ reaches the stationary response coefficient. 

\begin{figure}[t]
    \centering
    \includegraphics[width=7.5cm]{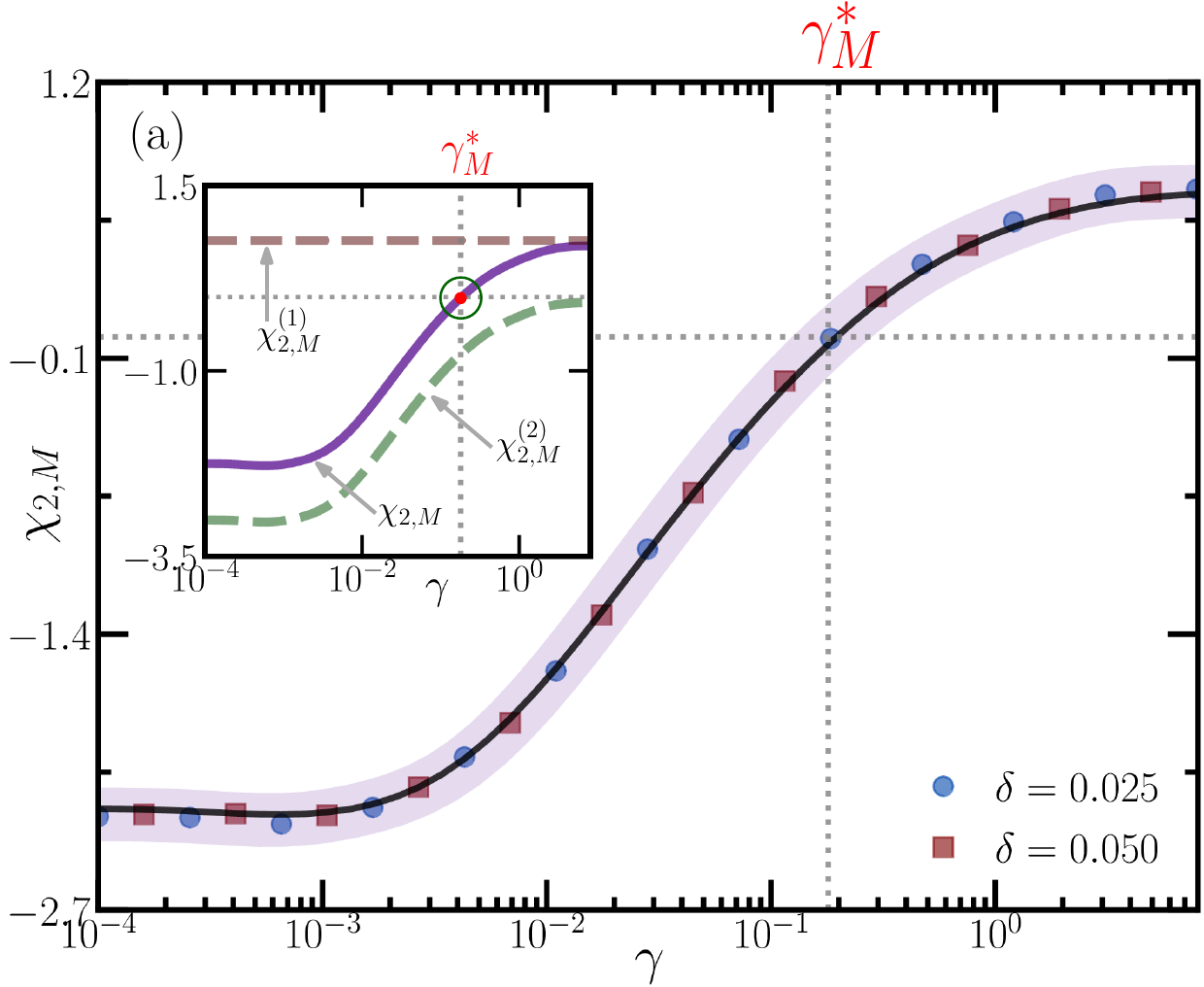}\includegraphics[width=7.5cm]{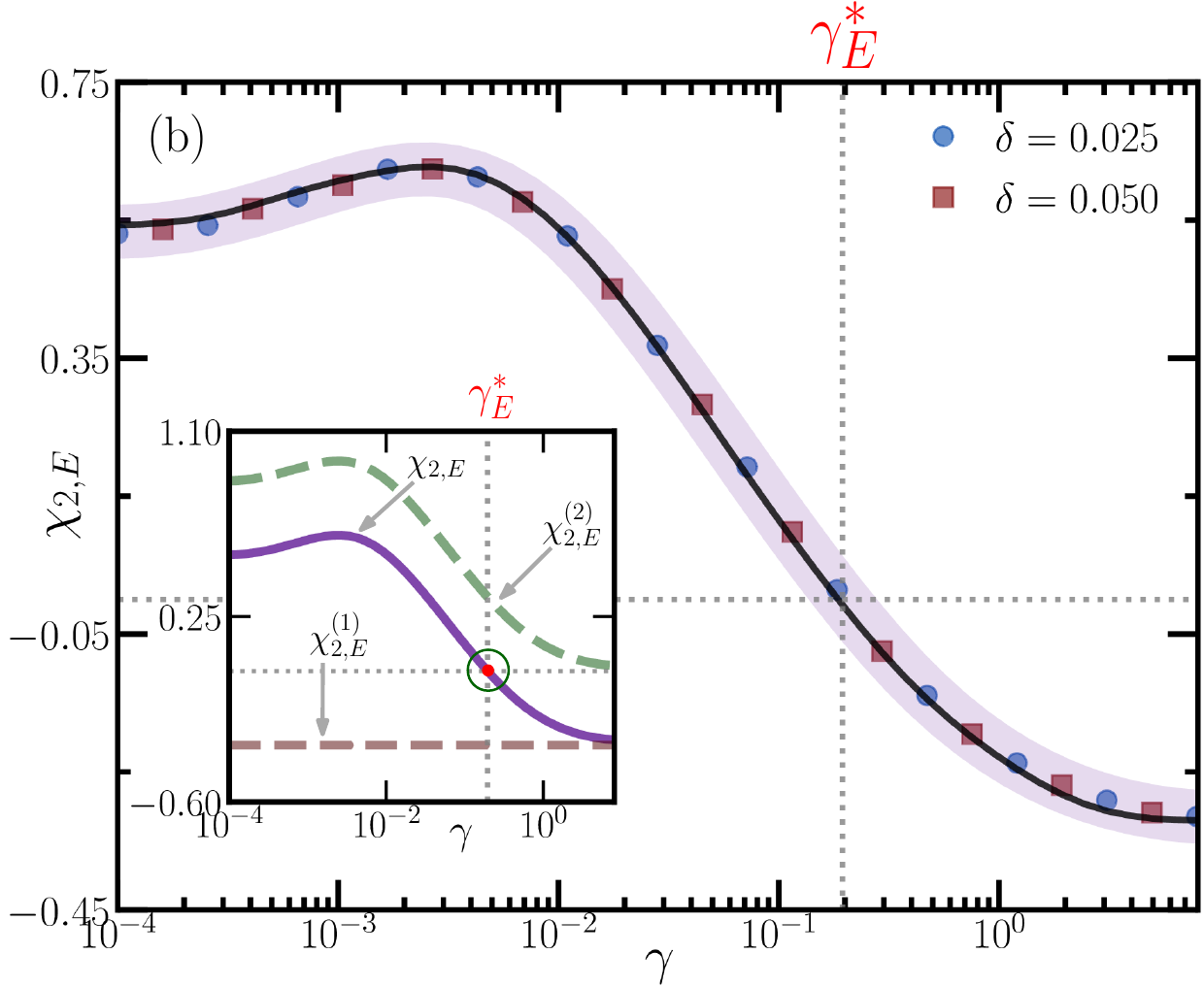}
    \caption{Small $\delta$ regime: Second order response coefficients for (a) average magnetization and (b) average energy, as functions of flip rate $\gamma$. The symbols correspond to the data obtained from numerical simulation for different values of $\delta$ whereas the black solid lines indicate the same obtained using response formalism [see \erefs{eq:chi2_break}-\eqref{eq:chi22}]. The shaded regions indicate the error margin corresponding to symbols. The inset shows the $\gamma$ dependent \big($\chi_{2}^{(2)}$\big) and independent \big($\chi_{2}^{(1)}$\big) components of the response coefficient separately. Here we have used $L=16$ and $\psi_0=0.1$.}
    \label{fig:Og_smdelT}
\end{figure}

It is noteworthy that the net energy gained $U(t)=H(t)-H(0)$, where $H(t)=-\sum_{\la ij\ra}s_{i}(t)s_{j}(t)$ denotes the total energy of the system at time $t$. Thus, the entropic component of the susceptibility \eref{eq:chi21} can be expressed as,
\begin{align}
    \chi_{2}^{(1)}(t) = -\frac{1}{T_{c}^{3}}\Big[\big\la H(t)O(t)\big\ra_{\rm eq}-\big\la H(0)O(t)\big\ra_{\rm eq}\Big].
\end{align}
In the limit $t\to\infty$, the correlation $\big\la H(0)O(t)\big\ra_{\rm eq}$ factorizes and $\chi_{2}^{(1)}$ can be expressed in the more familiar form of equilibrium response coefficient,
\begin{align}
    \chi_{2}^{(1)} = \frac{1}{T_{c}}\la V;O\ra_{\rm eq},\label{eq:kubo}
\end{align}
where $V=-\delta^{2} H / T_{c}^{2}$ can be interpreted as the perturbation to the original Hamilonian \eqref{eq:ham}.
Evidently the entropic term $\chi_{2}^{(1)}$ is independent of $\gamma$ and the $\gamma$ dependence of the susceptibility $\chi_{2}$ originates solely from $\chi_{2}^{(2)}$.

Using \erefs{eq:chi2_break}-\eqref{eq:chi22} we compute the stationary response coefficients $\chi_{2,M}$  and $\chi_{2,E}$ for magnetization and energy, respectively, from numerical simulations of the equilibrium Ising model at $T_{c}$. This is compared with the response measured directly from the perturbed system $\tilde{\chi}_{2,O}=\big(\la O \ra_\delta-\la O \ra_{\rm eq}\big)/\delta^{2}$ in Fig.~\ref{fig:Og_smdelT}, which shows an excellent match with the prediction \eref{eq:chi2_break}. From Fig.~\ref{fig:Og_smdelT}, it is clear  that the response coefficient vanishes at an intermediate value of $\gamma$, where,
\begin{align}
    \chi_{2}^{(2)}(\gamma^*_{O})=-\chi_{2}^{(1)}.
\end{align}
Thus, there exists a $\gamma^{*}_{O}$ where the expected value of the observable $\la O\ra$ in the NESS is equal to its equilibrium value. This explains the crossing of $\la M\ra_{\gamma}$ and $\la E\ra_{\gamma}$ for different values of $\delta$ shown in Figs.~\ref{fig:Mg}(a) and \ref{fig:Eg}(a). 

It is now useful to consider what happens to the susceptibility in the limiting situations $\gamma\to0$ and $\gamma\to\infty$. In the limit of vanishing switching rate $\gamma\to0$, the temperature remains constant at its initial value $T_{\sigma_{0}}$, allowing us to substitute $\sigma(t')=\sigma_{0}$ in \eref{eq:chi22}, which leads to,
\begin{align}
    \chi_{2}^{(2)}(t) \xrightarrow[]{\gamma\to 0}  -\frac{\psi_{0}}{2 T_{c}^{4}} 
    \sum_{\sigma_{0}} \rho(\sigma_{0}) \sigma_{0}^{2} 
    \Big\langle O(t) \, U(t) \int_{0}^{t} \! dt' \sum_{k=1}^{L^{2}} \Delta \varepsilon_{k}(t') \, 
    e^{ -\beta_{c} \varepsilon_{k}(t')/2 } \Big\rangle_{\rm eq}.\label{eq:chi2_g0}
\end{align}
Performing the sum over $\sigma_{0}$ and using \eref{eq:chi2_g0} in \eref{eq:master_response}, we get
\begin{align}
    \la O \ra_{\delta} \xrightarrow[]{\gamma\to0} \la O \ra_{\rm eq} -\frac{1}{T_{c}^{3}}\la O(t) U(t) \ra_{\rm eq} -\frac{\psi_{0}}{2T_{c}^4}\Big\la O(t) U(t) \intop_{0}^{t}dt'\sum_{k=1}^{L^{2}}\Delta \varepsilon_{k}(t') e^{-\beta_{c}\varepsilon_{k}(t')/ 2} \Big\ra_{\rm eq},\label{eq:O_sd_g0}
\end{align}
which infact is the small $\delta$ expansion of \eref{eq:sg_O} in the $\gamma\to0$ limit.

On the other hand, for large switching rate, $\gamma\gg\psi_{0}$, we find that the $\gamma$ dependent part of the response function $\chi_{2}^{(2)}\sim 1/\gamma$ [see Appendix~\ref{app:chi2} for details]. Hence, in the limit $\gamma\to \infty$, $\chi_{2}^{(2)}\to0$ and the susceptibility becomes solely entropic in nature [see \eref{eq:kubo}]. This is suggestive of an equililibrium-like picture in the large $\gamma$ scenario, which we investigate in the following section.

\section{\texorpdfstring{Large $\gamma$ regime and effective temperature description}{Large gamma regime and effective temperature description}}\label{sec:LG}
In this section, we focus on the behavior of the NESS in the large $\gamma$ regime, i.e., when $\gamma\gg\psi_{0}$. In this regime, the typical interval between two successive temperature switching events is much smaller than the typical time-scale of spin-flips. Hence, in the limit of infinite temperature switching rate, we can replace $\sigma$ by $\la \sigma \ra=0$ in \eref{eq:rate} to write the effective spin-flip rate as,
\begin{align}
    w_{\rm eff}(\Delta E) &\simeq \psi_0 \cosh\Big[\frac{\delta\beta_{\infty}\Delta E }{2T_c}\Big]\exp\Big[-\frac{\beta_{\infty}}{2}\Delta E\Big],~~~\text{with},~~~\beta_{\infty}^{-1}\equiv T_{\infty}=T_{c}\Big(1-\frac{\delta^{2}}{T_{c}^2}\Big).\label{eq:rate_gi}
\end{align}
Clearly, $w_{\rm eff}(\Delta E) / w_{\rm eff}(-\Delta E)=\exp(-\beta_{\infty} \Delta E)$ and \eref{eq:rate_gi} corresponds to a detailed balanced Glauber dynamics at an effective temperature $T_{\infty}$. Thus, in the limiting scenario of $\gamma\to\infty$, we expect the spin configuration weight to have the Boltzmann form $P(C)\sim e^{-\beta_{\infty}H(C)}$. Physically, this corresponds to a scenario where the temprature switches happens so fast that the Ising spins are not able to distinguish between the two reservoirs.

Nonequilibrium stationary states can sometimes be partially characterised by an effective temperature picture~\cite{Cugliandolo_2011}. For example, it has been shown that an Ising model in contact with two or more heat reservoirs  can sometimes be described using an equilibrium-like picture with an effective temperature~\cite{garrido1987stationary,Tamayo1994}. An effective temperature picture also describes a generalised FDT for nonequilibrium critical dynamics of Ising model~\cite{PhysRevE.68.016116}.

\begin{figure}[t]
    \centering
    \includegraphics[width=7.5cm]{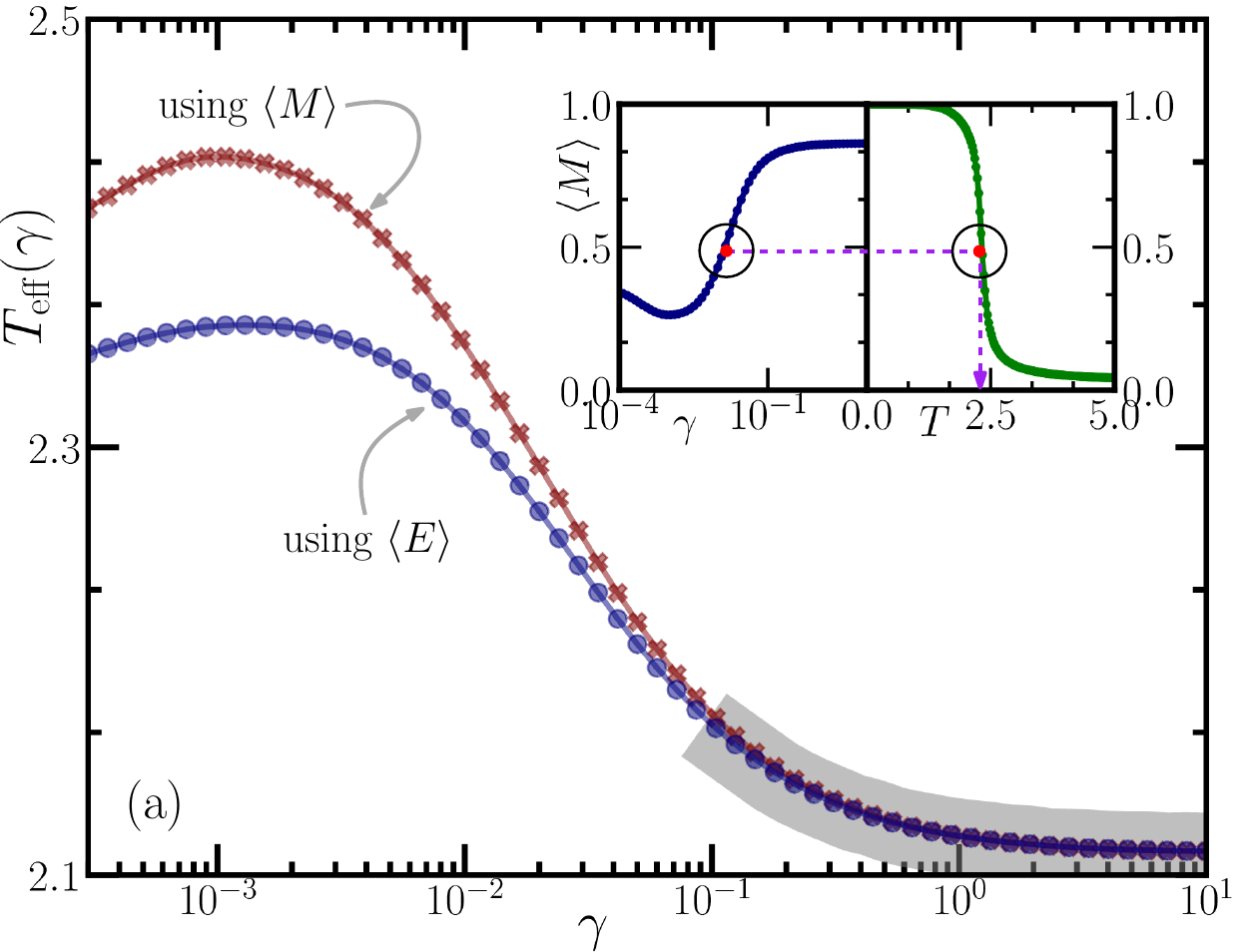}\includegraphics[width=7.5cm]{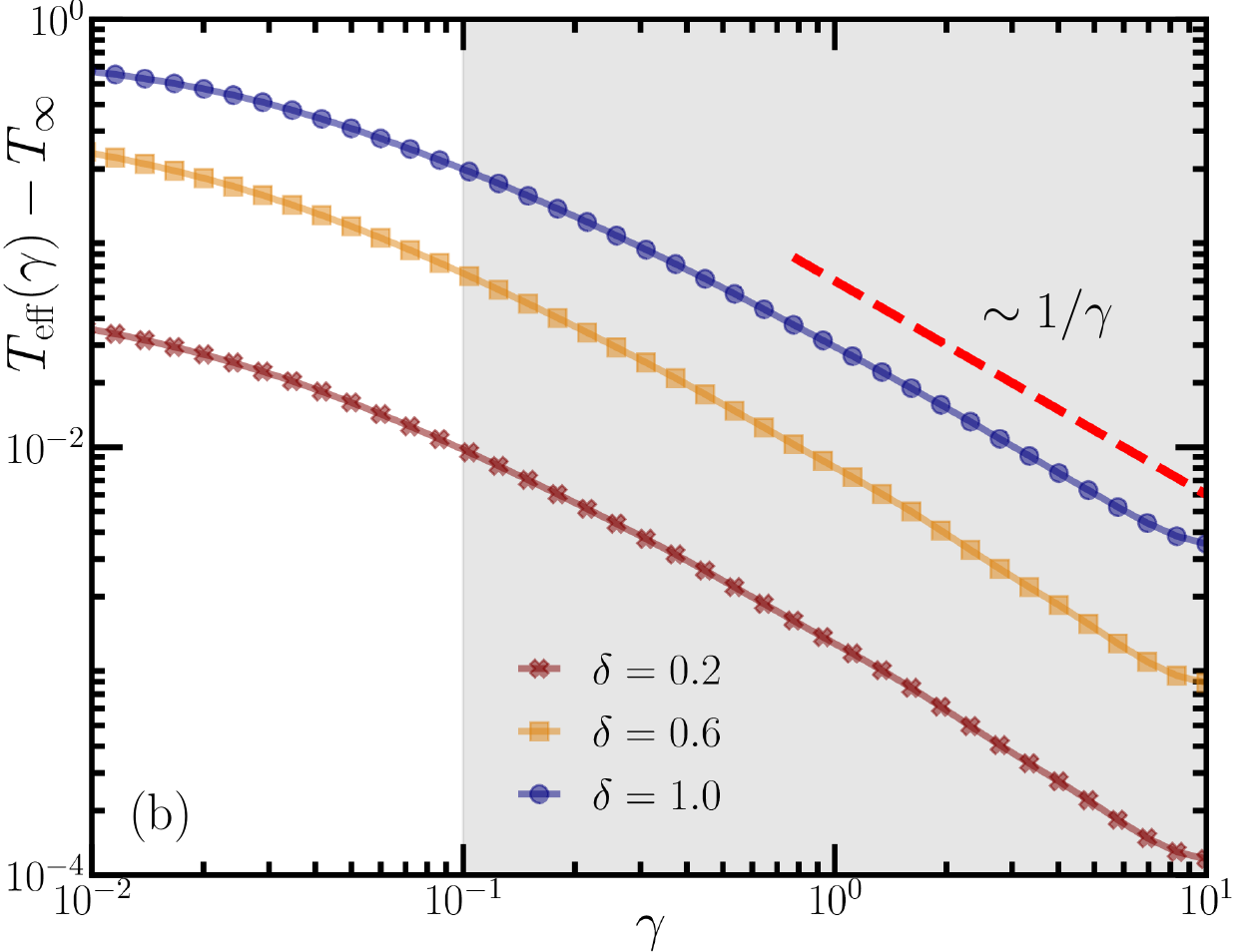}
    \caption{Large $\gamma$ regime: (a) Effective temperature $T_\text{eff}(\gamma)$, extracted from average magnetization and energy, as functions of the flip-rate $\gamma$, for a fixed value of $\delta = 0.6$. The gray shaded region indicates where the two curves coincide. The inset illustrates how $T_{\rm eff}$ is extracted from $\la M \ra$ vs $T$ curve for equilibrium Ising model [see \eref{eq:Teff_fromO_eq}]. (b) Plots of  $T_\text{eff}(\gamma)-T_{\infty}$ as functions of $\gamma$ for different values of $\delta$. The gray shaded region marks where $T_{\rm eff}(\gamma)$ is observable independent. Here we have used $L=32$ and $\psi_0=0.01$.}
    \label{fig:Tg}
\end{figure}

It is thus interesting to ask whether such an effective temperature picture also holds for the stochastic two-temperature Ising model for finite $\gamma$. In this case no such effective temperature can be directly obtained from the flip-rate. Instead, we extract an effective temperature $T_{\rm eff}(\gamma)$ by comparing $\la M \ra$ vs $\gamma$ curve with the equilibrium $\la M \ra$ vs T curve such that,
\begin{align}
    \la M \ra_{\gamma} = \la M \ra_{T_{\rm eff}(\gamma)}\label{eq:Teff_fromO_eq}
\end{align}
for a given value of $\delta$. This process is illustrated in the inset of Fig.~\ref{fig:Tg}(a). Similarly, a $\gamma$ and $\delta$ dependent effective temperature can also be extracted from $\la E \ra_{\gamma}$. Figure.~\ref{fig:Tg}(a) shows the plots of these two effective temperatures as functions of $\gamma$ for a particular value of $\delta$---clearly for $\gamma\gg\psi_{0}$, $T_{\rm eff}(\gamma)$ becomes independent of the choice of the observable, indicating the existence of a unique effective temperature. Infact, the effective temperatures extracted from $\la M^{2} \ra$ and $\la E^{2} \ra$ also correspond to the same values of $T_{\rm eff}(\gamma)$, although these are not explicitly shown here. Note that in the limit $\gamma \to\infty$, $T_{\rm eff}(\gamma)\to T_{\infty}$. Figure~\ref{fig:Tg}(b) illustrates that $T_{\rm eff}(\gamma)-T_{\infty}$ decays algebraically as $\sim 1/\gamma$ for different values of $\delta$.

\begin{figure}[t]
    \centering
    \includegraphics[width=8.5cm]{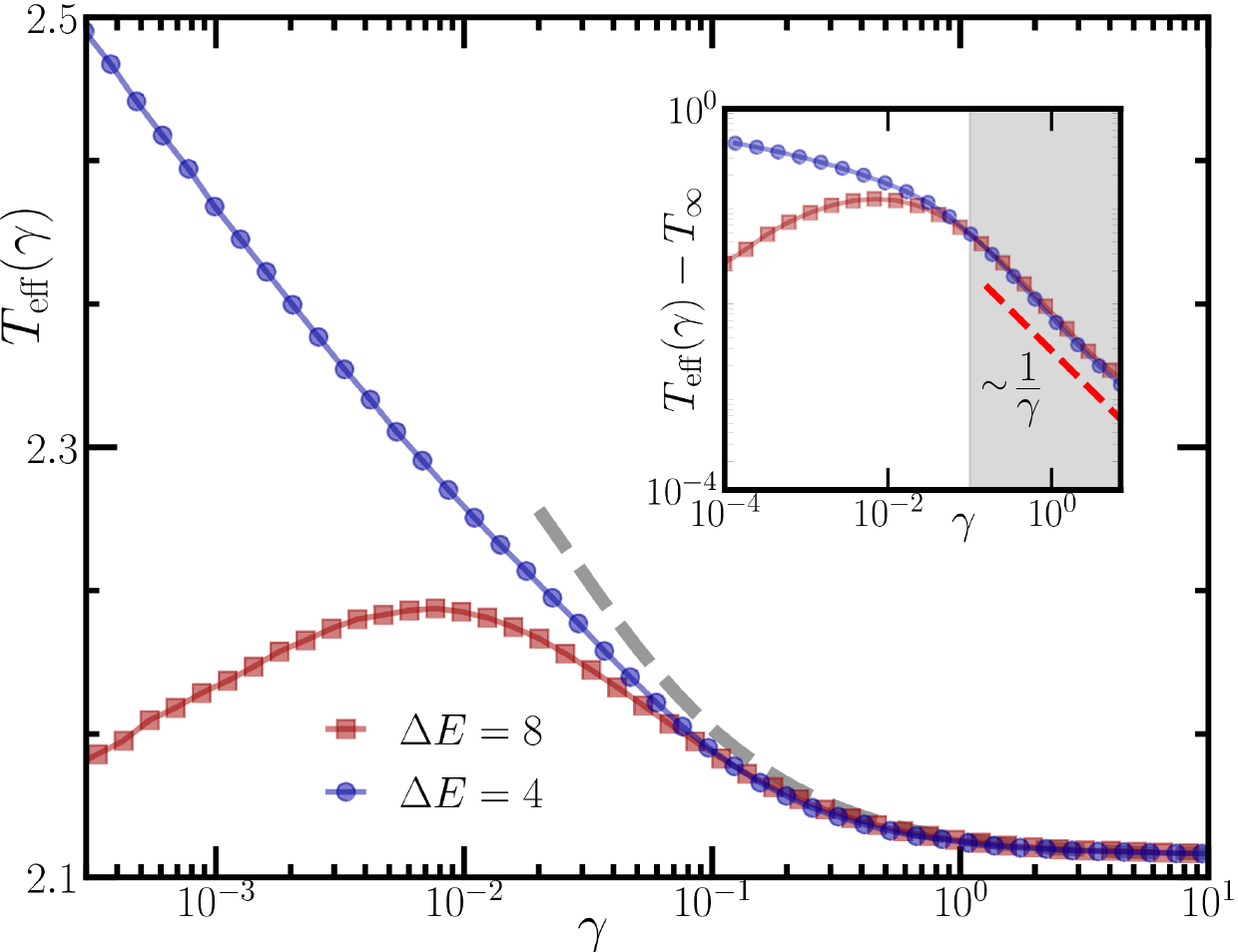}
    \caption{Effective temperature $T_{\rm eff}(\gamma)$ as function of flip-rate $\gamma$, extracted from probability of spin flips corresponding to energy change $\Delta E=4$ and $\Delta E = 8$. The gray dashed line indicates $T_{\rm eff}(\gamma)$ extracted from $\la E \ra$ vs $\gamma$ curve [see \eref{eq:Teff_fromO_eq} and Fig.~\ref{fig:Tg}(a)]. The inset shows  $T_{\rm eff}(\gamma) - T_{\infty}$ decays as $1/\gamma$. The gray shaded region marks where $T_{\rm eff}(\gamma)$ is observable independent. Here we have used $L=32$, $\psi_{0}=0.01$ and $\delta=0.6$.}
    \label{fig:Tg_dynamics}
\end{figure}

Such an effective temperature description is not expected to hold when the time-scales of temperature switch and spin-flip become comparable. This is also apparent from Fig.~\ref{fig:Tg}(a), which illustrates that the $T_{\rm eff}(\gamma)$ curves obtained from $\la M \ra_{\gamma}$ and $\la E \ra_{\gamma}$ no longer match, in the small $\gamma$ regime.

The existence of a unique effective temperature suggests that the weight of the spin configuration $C$ in the NESS admits a Boltzmann-like form $P(C) \sim \exp \big[- H(C)/ T_\text{eff}\big]$ in the large $\gamma$ regime. Such a Boltzmann-like form for the staionary state weight however does not necessarily imply that the system is in equilibrium~\cite{katz1983phase,Luck_2006,de2017gibbsian}. To understand whether the spin degrees indeed achieves equilibrium, we measure the effective spin-flip rates.

For fixed $\gamma$ and $\delta$, the effective spin-flip rate $w_{\rm eff}(\Delta E)$ can be estimated from the ratio of number of successful spin-flips to attempted spin-flips with energy change $\Delta E$. From these rates, an effective temperature can be extracted as,
\begin{align}
    T_{\rm eff}&=-\frac{\Delta E}{\log\Big[w_{\rm eff}(\Delta E)\big/w_{\rm eff}(-\Delta E)\Big]}
\end{align}
From numerical simulations, we extract $T_{\rm eff}$ for 
the allowed values of $\Delta E=\pm 4$ and $\pm 8$. These are compared with the $T_{\rm eff}(\gamma)$ obtained from the stationary values of the observables using \eref{eq:Teff_fromO_eq} [see Fig.~\ref{fig:Tg_dynamics}]. Remarkably, for $\gamma\gg\psi_{0}$, all the curves converge, indicating that the dynamics effectively does satisfy detailed balance. Thus it appears that the spin degrees, when considered independently of the bath degrees of freedom, exhibit an effective equilibrium-like behaviour in this large switching rate regime.

\begin{figure}
    \centering
    \includegraphics[width=7.5cm]{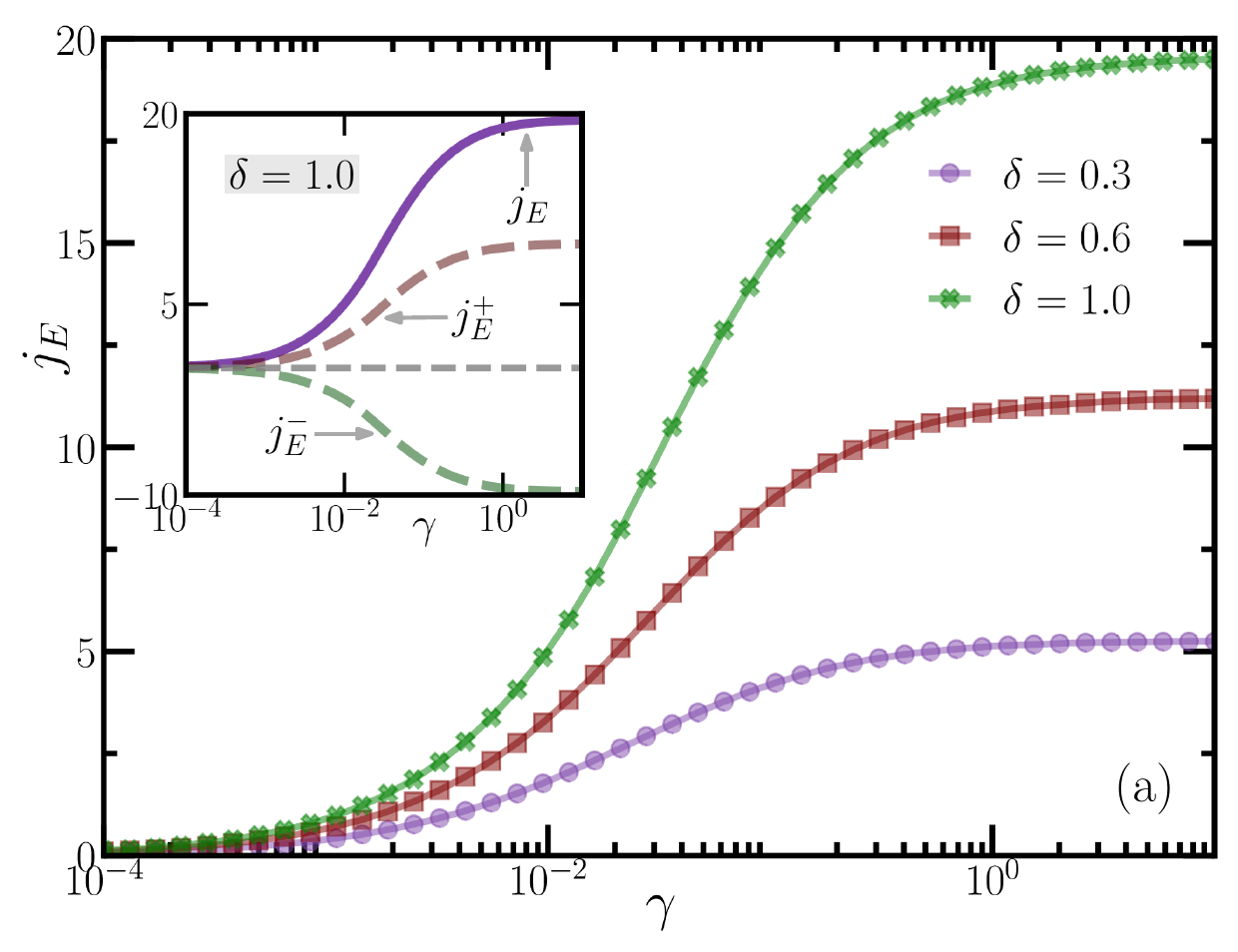}\includegraphics[width=7.5cm]{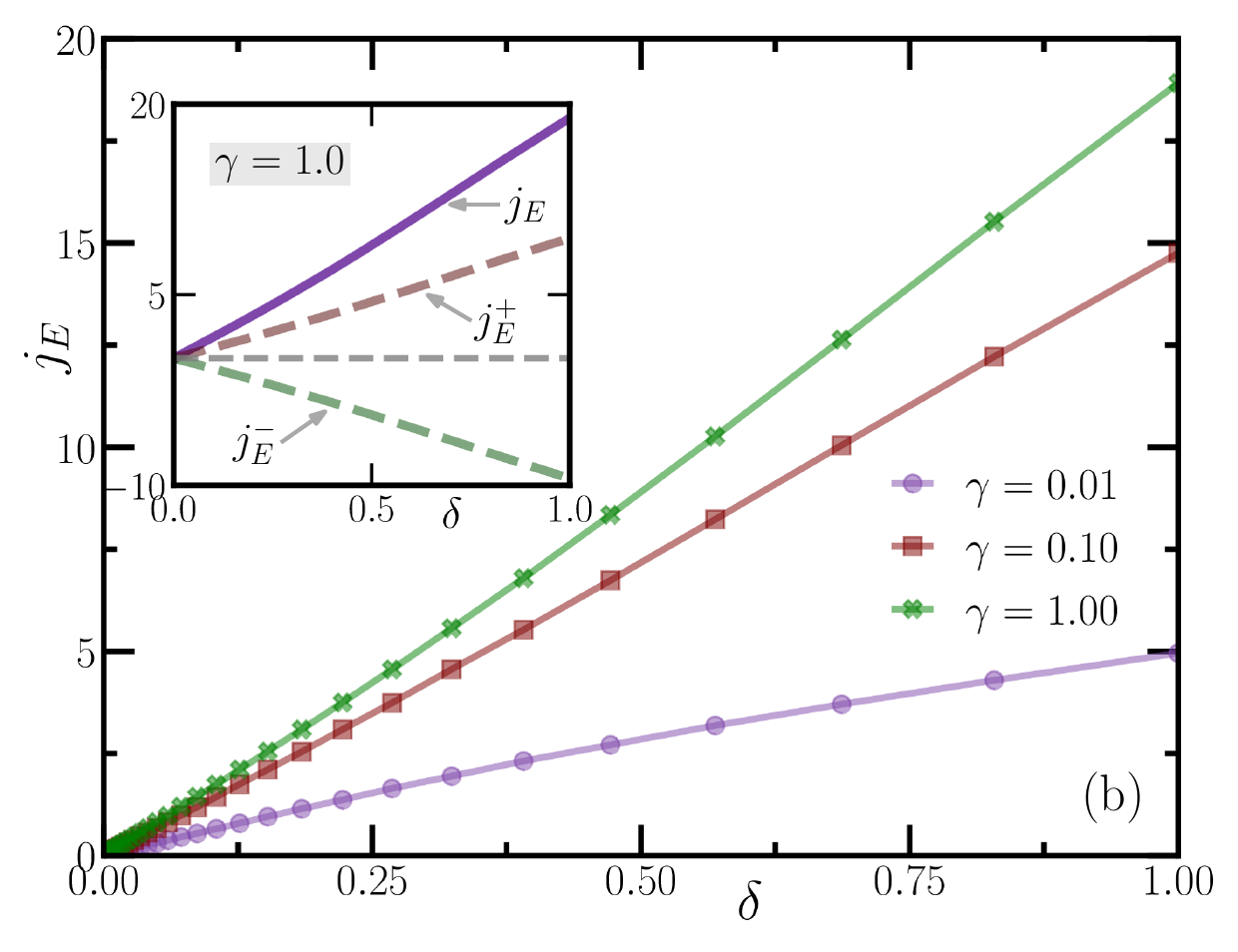}
    \caption{Energy current:  Plots of (a) $j_{E}$ versus $\gamma$ for different values of $\delta$, and (b) $j_{E}$ versus $\delta$ for different values of $\gamma$. The insets separately show the average energy currents $j_{E}^{\pm}$ flowing from the hot and the cold reservoirs into the system. Here we have used $L=32$ and $\psi_{0}=0.01$.}
    \label{fig:jg_jdelT}
\end{figure}

However, the full system consisting of the spin degrees of freedom and the two reservoirs is inherently nonequilibrium in nature. To illustrate this, we measure dynamical observables like the energy current flowing through the system, which should vanish in equilibrium. The average instantaneous current flowing to the system from the reservoir at temperature $T_{\sigma}$ in the stationary state is given by,
\begin{align}
    j_{E}^{\sigma}&=\lim_{t\to \infty} \frac 1t \la \Delta E^{\sigma}(t)\ra,
\end{align}
where $\Delta E^{\sigma}$, introduced in \erefs{eq:Sp_eq}-\eqref{eq:Spp_eq}, is the total energy gained by system from the reservoir during the time interval $[0,t]$. Then the average instantaneous energy current flowing through the system from the hot reservoir to the cold reservoir is given by,
\begin{align}
    j_{E}&=j_{E}^{+}-j_{E}^{-}
\end{align}
Figure~\ref{fig:jg_jdelT}(a) shows the plot for $j_{E}$, obtained from numerical simulations, as a function of the switching rate $\gamma$ for different value of $\delta$. Evidently, $j_{E}$ increases with increasing $\gamma$ and eventually saturates. The average current also increases with increasing $\delta$ as shown in Fig.~\ref{fig:jg_jdelT}(b). Thus a finite energy current always flows from the hot reservoir to the cold reservoir, notwithstanding the effective temperature description in the large $\gamma$ regime. This is a signature of the inherent nonequilibrium nature of the stochcastic two-temperature Ising model dynamics.

\section{Finite-size Effects}

\begin{figure}
    \centering
    \includegraphics[width=7.5cm]{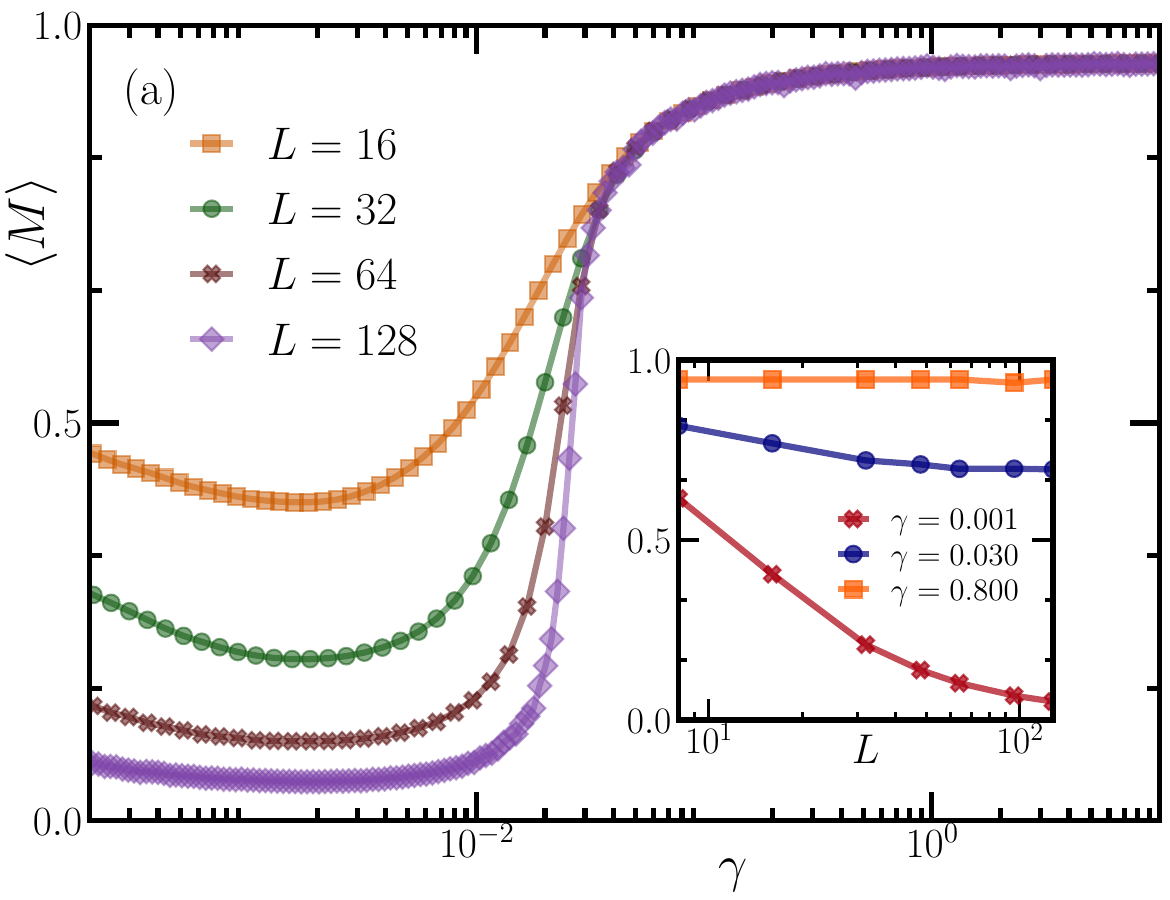}\includegraphics[width=7.5cm]{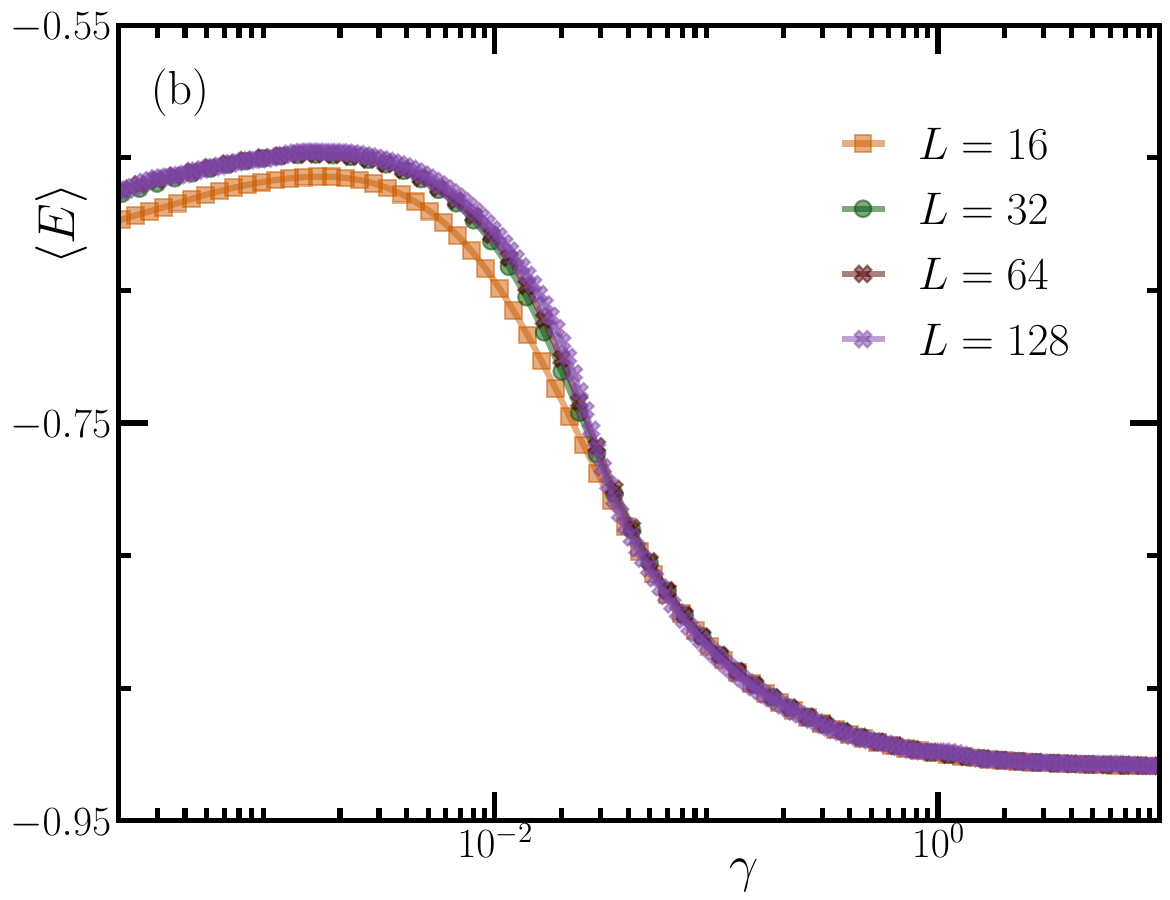}
    \caption{Finite size effect: (a) Average magnetization $\la M\ra$ and (b) average energy $\la E \ra$ as functions of flip-rate $\gamma$ for different system sizes $L$. The inset in panel (a) shows plots of $\la M\ra$ versus $L$ for different values of $\gamma$. Here we have used $\psi_{0}=0.01$ and $\delta=1.0$.}
    \label{fig:L_dep}
\end{figure}

In the previous sections, we have analysed the behaviour of the NESS using various theoretical techniques and limiting scenarios that are independent of the system size. However, so far we have relied on a fixed system size to validate these results. It is therefore natural to examine how the observables considered depend on the system size. To this end, we investigate the system-size dependence of the average magnetization and the energy in the stationary state. Figure~\ref{fig:L_dep} shows plots of $\la M \ra$ and $\la E \ra$ as functions of $\gamma$ for different values of system size $L$. Clearly, the non-monotonic nature of both the average magnetization and energy remain robust across system size variation. In fact, the qualitative behaviour of both the observables remain same as the system size is varied, although the magnetization shows a pronounced quantitative dependence on $L$.

The inset in Figure~\ref{fig:L_dep}(a) shows the behaviour of $\la M \ra$ as a function of the system size $L$ for different values of $\gamma$. Clearly, for small $\gamma \ll \psi_0$, the magnetization decreases with increasing $L$, while for large $\gamma$, it becomes largely independent of $L$. This can be understood as follows. As already discussed, in the $\gamma \to 0$ limit [see \eref{eq:sg_O}], the magnetization is given by,
$\la M \ra_\gamma = [\la M \ra_\text{eq}^+ + \la M \ra_\text{eq}^-]/2$, where $\la M \ra_\text{eq}^\sigma$ denotes average magnetization of equilibrium Ising model at temperature $T_\sigma$. In the limit of thermodynamically large system size, $\la M \ra_\text{eq}^+ \to 0$ while $\la M \ra_\text{eq}^-$ attains a finite value, leading to $\la M \ra_\gamma \simeq \la M \ra_\text{eq}^-/2$. On the other hand, as discussed in Sec~\ref{sec:LG}, for $\gamma \to \infty$, the spin-configurations in the NESS has an effective equilibrium description at a temperature $T_\text{eff} < T_c$. This 
implies that for large $\gamma$, the system attains an ordered state, yielding a large value of the magnetization. The average energy shows an opposite trend---for small $\gamma$, $\la E \ra$ increases as $L$ is increased. Moreover, the overall system size dependence of the average energy is less pronounced, as can be seen from Fig.~\ref{fig:L_dep}(b).

It should be noted that both $\langle M \rangle$ and $\langle E \rangle$ exhibit little dependence on the system size in the large-$\gamma$ regime, implying that the effective temperature at large $\gamma$ is essentially independent of the system size.

\section{Conclusion}\label{sec:conc}
In this work, we have studied the two-dimensional Ising model subjected to a stochastic dichotomous temperature. The temperature alternates between $T_c \pm \delta$ with a rate $\gamma$ which leads the system to a nonequilibrium stationary state characterised by unusual behaviour of observables like magnetization and energy. For example, the average magnetization and average energy change non-monotonically as functions of flip-rate $\gamma$. We have characterised this NESS through two complementary approaches. In the small $\gamma$ regime, we use a renewal approach to explain the origin of the non-monotonic variation of the obseravbles. On the other hand, for small $\delta$, we employ the dynamical response theory to understand the $\gamma$ dependence of the observables. 

We find that the NESS admits an effective temperature description for fast temperature switches, i.e., for large values of $\gamma$. We explicitly compute this effective temperature in the $\gamma \to \infty$ limit, which turns out to be below the critical temperature of the ordinary Ising model. Using extensive numerical simulation, we show that the effective temperature description remains valid for a finite range of $\gamma$ and breaks only in the small $\gamma$ regime. 
We further show that such an effective temperature description does not imply that the system reaches an equilibrium state---the inherent nonequilibrium nature of the dynamics leads to a non-zero energy current flowing through the system from the hot reservoir to the cold reservoir. We also show that the qualitative behaviour of the observables in the NESS remains robust across system size variation.

A natural open question is whether similar effects are observed in other systems that belong to the Ising universality class. In a more general sense, our study can be thought of as a special case of a driving protocol where a control parameter of a system is randomly switched between two distinct values according to a dichotomous process. In this context, it would be worthwhile to investigate systems with underlying nonequilibrium dynamics like the Katz-Lebowitz-Spohn model~\cite{katz1983phase} or the driven lattice gas~\cite{schmittmann1995statistical}. 
It would also be of interest to explore whether our results can be tested experimentally using thin-film set-ups similar to the ones used in Ref.~\cite{expt_1,adma_202501043}.

\section*{Acknowledgements}
We acknowledge the National Supercomputing Mission (NSM) for providing computing resources of ‘PARAM RUDRA’ at S.N. Bose National Centre for Basic Sciences, which is implemented by C-DAC and supported by the Ministry of Electronics and Information Technology (MeitY) and Department of Science and Technology (DST), Government of India.

\vspace{0.25cm}

\noindent {\bf Funding information:} RS acknowledges the CSIR grant no. 09/0575(11358)/2021-EMR-I. UB acknowledges the support from the Anusandhan National Research Foundation (ANRF), India, under a MATRICS grant [no. MTR/2023/000392].

\section*{Data Avaiability}
The data that support the findings of this article are obtained from numerical simulations. These are available from the authors upon reasonable request.

\appendix
\numberwithin{equation}{section}

\section{Response coefficient in the limiting case of infinite switching rate}\label{app:chi2}
In this appendix, we derive the expressions for second-order response coefficient $\chi_{2}(t)$ in the limiting scenario of infinite switching rate, i.e., for $\gamma\to\infty$. We start by considering the $\gamma$ dependent part of the response coefficient quoted in \eref{eq:chi22},
\begin{align}
    \chi_2^{(2)}(t) = -\frac{\psi_{0}}{2T_c^4}\Big\langle O(t)\intop_{0}^{t}dt''\sigma(t'')j_{E}(t'')\intop_{0}^{t}dt'~\sigma(t')\sum_{k=1}^{L^{2}}\Delta \varepsilon_{k}(t')\,e^{-\beta_{c}\Delta \varepsilon_{k}(t')/2}\Big\rangle_{\rm eq}.\label{eq:chi22_sim0}
\end{align}
In writing the above expression we have used,
\begin{align}
    \mathscr{J}_{E}(t)=\intop_{0}^{t}dt''\sigma(t'')\mathscr{I}(t''),
\end{align} 
in \eref{eq:chi22}, where $\mathscr{I}(t)$ denotes the instantaneous energy current flowing into the system. In the unperturbed dynamics, i.e., for $\delta=0$, the dichotomous process $\sigma\to-\sigma$ and the Ising spin-flip dynamics at temperature $T_{c}$ become statistically independent, allowing the averages over $\sigma$ and over the spin configurations to be factorised. Consequently, \eref{eq:chi22_sim0} now simplifies to
\begin{align}
    \chi_2^{(2)}(t) = -\frac{\psi_{0}}{2T_c^4}\intop_{0}^{t}dt'\intop_{0}^{t}dt''~\big\la\sigma(t')\sigma(t'')\big\ra_{\rm eq}\Big\langle O(t)j_{E}(t'')\sum_{k=1}^{L^{2}}\Delta \varepsilon_{k}(t')\,e^{-\beta_{c}\Delta \varepsilon_{k}(t')/2}\Big\rangle_{\rm eq}.\label{eq:chi22_sim1}
\end{align}
For large $\gamma$, $\sigma$ effectively emulates a white noise with autocorrelation,
\begin{align}
\la \sigma(t')\sigma(t'')\ra\to\frac{1}{\gamma}\delta(t'-t'').\label{eq:autocorr_gi}
\end{align}
Substituting \eref{eq:autocorr_gi} in \eref{eq:chi22_sim1} and subsequently performing the integral over $t''$, yields
\begin{align}
    \chi_2^{(2)}(t) = -\frac{\psi_{0}}{2\gamma T_c^4}\intop_{0}^{t}d t' \Big\langle O(t)j_{E}(t')\sum_{k=1}^{L^{2}}\Delta \varepsilon_{k}(t')\,e^{-\beta_{c}\Delta \varepsilon_{k}(t')/2}\Big\rangle_{\rm eq}.\label{eq:chi22_sim2}
\end{align}
It is thus evident from the above expresion that in the limit $\gamma\to\infty$, $\chi_{2}^{(2)}$ vanishes like $1/\gamma$.

\section*{Bibliography}
\bibliographystyle{iopart-num}
\bibliography{fising}
\end{document}